\documentclass[11pt, executivepaper]{article}
\usepackage[utf8]{inputenc}
\usepackage[T1]{fontenc}
\usepackage{natbib}
\usepackage{amsmath}
\usepackage{mathtools}
\usepackage{xcolor}
\usepackage{amsfonts}
\usepackage{graphicx}
\usepackage{enumitem}
\usepackage{geometry}
\usepackage{hyperref}
\hypersetup{colorlinks= true, allcolors=blue}
\setcitestyle{aysep={}}

\begin{document}

\title{\textbf{Hidden Variables in Quantum Mechanics: A Journey Into de Broglie-Bohm Pilot-Wave Theory}}

\author{Andrea Oldofredi\thanks{Contact Information: University of Lisbon, School of Arts and Humanities, Centre of Philosophy. E-mail: aoldofredi@letras.ulisboa.pt}}

\maketitle

\begin{abstract}

This work provides an introduction to the most influential hidden-variable interpretation of the quantum formalism, i.e., the de Broglie-Bohm pilot-wave theory. As the reader will see, the chapter offers a detailed discussion of the main formulations of the theory. Moreover, cogent topics such as the origin of the Born rule, the interpretation of probability, the nature of the wave function, the status of spin, as well as extensions of the pilot-wave approach beyond non-relativistic quantum mechanics will be examined. This essay, however, not only reviews existing relevant literature but also contributes new insights about (i) the existence of spin in the context of pilot-wave theory, and (ii) whether the latter framework is formally and conceptually simpler than standard quantum mechanics. 
\vspace{5mm}

\vspace{5mm}

\noindent \emph{Keypoints}: 
\begin{itemize}
\item The paper shows that Pilot-Wave Theory is structurally simpler than Quantum Mechanics;
\item The paper provides arguments for the existence of spin in Pilot-Wave Theory;
\item The paper reviews the main formulations, philosophical issues, and extensions of Pilot-Wave Theory.
\end{itemize}

\vspace{5mm}

\noindent  Forthcoming in \emph{Comprehensive Philosophy of Science}, Chapter 7.33, edited by S.O. Hanson, Section Editor A. Wilson, Elsevier.
\vspace{4mm}

\end{abstract}

\clearpage 
\tableofcontents
\newpage

\section{Introduction: Hidden Variables and the Completeness of Quantum Mechanics}
\label{Intro}

The possibility of completing Quantum Mechanics (QM) with hidden variables is one of the oldest foundational questions discussed by philosophers and physicists since the theory's inception---together with the issue of its ontological intelligibility, or {\em Anschaulichkeit} (cf.\ \cite{Regt:1997, Regt:2017}). Whereas many of the founding figures of QM held that it provides the maximal information about the state of physical systems, eminent physicists cast doubts about this belief.\footnote{For detailed historical discussions cf.\ \cite{Bacciagaluppi:2009}, \cite{Jammer:1974aa}, \cite{Seth:2013}.} 

Referring to this, the celebrated and widely discussed Einstein-Podolsky-Rosen (EPR) paper was published in 1935, effectively marking the beginning of discussions concerning the statistical nature of quantum theory---a longstanding position advocated mainly by Einstein---and its incompleteness (\cite{EPR:1935}).\footnote{For historical accuracy, it should be underlined that already in 1927 at the fifth Solvay conference, Einstein provided arguments against the completeness of quantum theory. See \cite{Bacciagaluppi:2009}, Chapter 7, for discussion.} In this essay, the authors maintained that, to be considered satisfactory, any physical theory should be empirically adequate and complete in its description of physical phenomena. As is well known, they aimed to show that QM is unsatisfactory because it is incomplete. Their argument is based on the following assumptions:
\begin{itemize}
\item {\bf Completeness:} Every element of the physical reality must have a counterpart in the physical theory;
\item {\bf Criterion of Reality:} If, without in any way disturbing a system, we can predict with certainty (i.e., with probability equal to unity) the value of a physical quantity, then there exists an element of physical reality corresponding to this physical quantity;
\item {\bf Locality:} Physical processes and events occurring at a certain spacetime region have no instantaneous influence or effect on other systems located at spacelike separated regions.\footnote{Contrary to the other two principles, which are explicitly stated by EPR, {\bf Locality} is implicitly assumed.}
\end{itemize}

\noindent To illustrate the EPR argument straightforwardly, let us consider an electron-positron pair that is emitted by a common source, as shown in Figure 1 below.\footnote{Here I offer a streamlined version of the EPR argument based on Bohm's presentation (\cite{Bohm:1951}).} The electron is sent in the right direction, where agent Alice is located, whereas the positron is sent in the left direction, where observer Bob is located. For the sake of the argument, let's suppose that a source placed between Alice's and Bob's locations prepares the pair in a singlet state---thus, the particles are entangled:
\begin{align}
\label{singlet}
|\psi\rangle= \frac{1}{\sqrt{2}} \bigg(|\uparrow\rangle_e|\downarrow\rangle_p - |\downarrow\rangle_e|\uparrow\rangle_p\bigg).
\end{align}

\noindent The above equation represents a superposition of two 2-particle states: in the first, the electron has $z$-spin-up, whereas the positron has $z$-spin-down; conversely, in the second state, the electron has $z$-spin-down, and the positron $z$-spin-up.\footnote{This choice is arbitrary; Alice and Bob may indeed decide to measure, e.g., $\hat{S}_y$  or $\hat{S}_x$.} 
\begin{center}
\includegraphics[scale=.60]{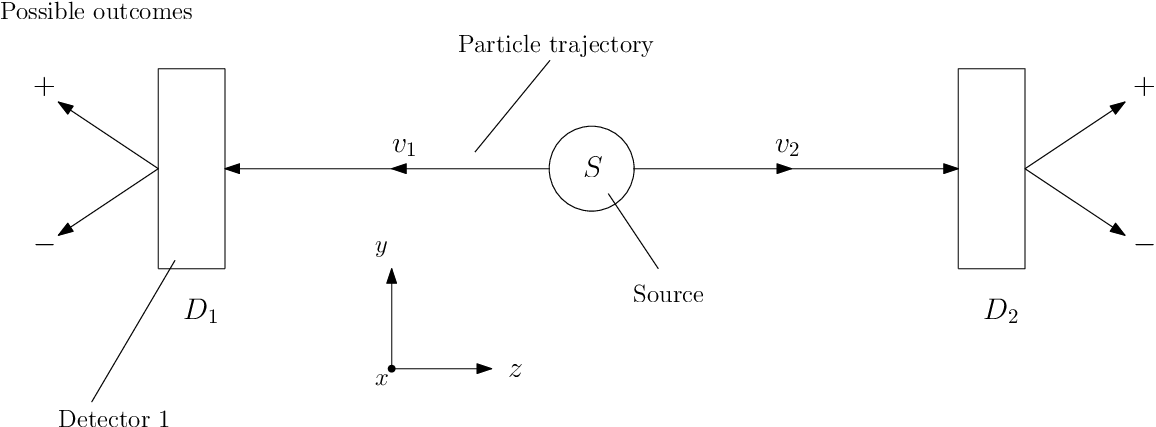}
\begin{quote}
\center \scriptsize{Figure 1: Idealized representation of the EPR-Bohm {\em Gedankenexperiment.}}
\end{quote}
\end{center}

\noindent The nature of the paradox emerges from the measurement of $\hat{S}_z$ on the electron (or positron), which allows us to infer with certainty the value of $\hat{S}_z$ for the position (or electron). For instance, if Alice measures the $z$-spin on the electron and finds, e.g., $z$-spin-up, we can predict with certainty the value of $\hat{S}_z$ for the positron without measuring it, for $\hat{S_z}_{B}=-\hat{S_z}_{A}$. Alice's measurement, then, causes a collapse of the total quantum state into one of the two possible superposed states, thereby determining the values of $\hat{S}_z$ for {\em both} particles. However, these corpuscles are space-like separated: {\bf Locality} then forbids that a measurement performed by Alice should affect the state of the particle at Bob's location. Therefore, either locality is violated, or the information given by the quantum mechanical wave function \eqref{singlet} is not complete, for there must be a physical element of reality---the value of the positron's spin---that is not included in the wave function describing the total state of the system. Consequently, given that for Einstein, Podolsky, and Rosen, locality was an essential principle of physics, they concluded that QM must be incomplete (cf.\ \cite{Norsen:2006}).

The EPR paper is then concluded with the following remark:
\begin{quote}
While we have thus shown that the wave function does not provide a complete description of the physical reality, we left open the question of whether or not such a description exists. We believe, however, that such a theory is possible (\cite{EPR:1935}, p.\ 780).
\end{quote}

\noindent This statement implies that QM may be completed with additional parameters; these are the so-called {\em hidden variables}. Consequently, for EPR, a deeper, complete framework of quantum mechanics integrated with such hidden parameters is a concrete possibility.

Contrary to this conclusion, however, there exist formal results that undermine such a belief. As is well known, already in von Neumann's first axiomatic presentation of QM in 1932, contained in his {\em Mathematische Grundlagen der Quantenmechanik}, one finds the first no-go theorem against the possibility of completing QM with additional parameters (\cite{vonNeumann:1932}). Precisely, von Neumann shows that it is not possible to recover 

\begin{quote}
the quantum probabilities from a hidden variable theory of dispersion-free\footnote{Dispersion-free states are those states in which the values of {\em every} observable are determined at the same time.} (deterministic) states in which the quantum observables are represented as the ‘beables’ of the theory, to use Bell’s suggestive term. That is, the quantum probabilities could not reflect the distribution of pre-measurement values of beables, but would have to be derived in some other way, e.g., as in Bohm's theory, where the probabilities are an artefact of a dynamical process that is not in fact a measurement of any beable of the system (\cite{Bub:2010}). 
\end{quote}

\noindent Alternatively, von Neumann answered in the negative to the question as to whether quantum systems could be completed with parameters in addition to the information contained in the quantum state, i.e., whether the specification of the values of such hidden parameters in conjunction with the quantum state could give the values of all physical quantities ``exactly and with certainty'' (von Neumann quoted in {\em ibid.}, p. 1335).\footnote{Von Neumann's theorem has been subject to severe criticisms. John Bell famously argued that the theorem rests on this assumption: ``Any real linear combination of any two Hermitian operators represents an observable, and the same linear combination of expectation values is the expectation value of the combination'' (\cite{Bell:2004aa}, p.\ 4). Bell argues that it implicitly assigns additive expectation values to noncommuting operators as well, and thus, it entails an impossible relation between the results of incompatible measurements. Similarly, Grete Hermann made the same objection already in 1933, cf.\ for details \cite{Hermann:1933}, \cite{Crull:2017}. \cite{Bub:2010} rebuts Bell-type criticisms and highlights the positive content of the theorem, which, as Bub claims, does not exclude {\em all} classes of hidden-variables theories. Cf. also \cite{Dieks:2017} for a critical assessment of Bell and Hermann's arguments against von Neumann's theorem.} 

In addition to von Neumann's theorem, other formal results imposed requirements and constraints on what kind of hidden variable theories are allowed in QM. On the one hand, Bell's theorem (\cite{Bell:2004aa}, Chapter 2) is interpreted as showing the non-local character of quantum mechanics; a hidden variable theory that can be empirically adequate must be nonlocal, contrary to EPR's hope to retain locality in QM. On the other hand, the Kochen-Specker theorem (\cite{Kochen:1967aa}) imposes that, contrary to the classical case, quantum observables are {\em contextual}, i.e., the value assigned to an observable cannot generally be specified independently of the measurement context, namely the set of compatible observables jointly measured. This formal requirement thereby prevents the possibility of assigning hidden variables specifying the values of a set of properties ascribed to a certain quantum system.

Notwithstanding these constraints, several hidden variable theories have been proposed over the years. Primary examples are Gerard 't Hooft's cellular automaton interpretation (\cite{Hooft:2016}), as well as Tim Palmer and Sabine Hossenfelder's works on Superdeterminism (cf., \cite{Hossenfelder:2020}). These approaches provide a {\em local} formulation of hidden variable theories, for they maintain a different interpretation of Bell's theorem (\cite{Hance:2022}).\footnote{Critical examinations of Superdeterminism are given in \cite{Chen:2022} and \cite{Baas:2023}. Moreover, {\em contra} Hance and Hossenfelder, \cite{Goldstein2011} and \cite{Maudlin:2014} claim that Bell's theorem shows that nature is fundamentally non-local.} 

This Chapter, however, focuses on the most influential formulation of hidden variable theory, namely the Pilot-Wave Theory (PWT). The essay is structured as follows: in Section \ref{PWT}, the main formulations of PWT are introduced in chronological order, from Louis de Broglie's initial proposal to the contemporary framework of Bohmian Mechanics (BM), passing through David Bohm's 1952 rediscovery of de Broglie's theory. Section \ref{philo} analyses the role and status of Born's rule in PWT, together with a philosophical discussion on the status of the wave function, which plays a pivotal role in the context of any PWT. Section \ref{measure} not only reviews the basic elements of the pilot-wave approach to quantum measurements, illustrating the meaning of quantum observables in this interpretation of QM and its contextual nature, but also argues that the structure of PWT is simpler with respect to that of standard QM, contrary to the general wisdom. Furthermore, Section \ref{Spin} contains the second and major original contribution of this paper. Here we provide the classical arguments used by supporters of BM to argue for the non-existence of spin, discussing as well some major physical consequences of this fact for the ground state of the Hydrogen atom. In this section, I suggest a new argument for the existence of spin in BM. In Section \ref{Extensions}, extensions of the PWT to quantum field theory and relativity are analyzed. Particular attention will be devoted to those proposals not sufficiently discussed in the philosophical literature, such as the Hypersurface Bohm-Dirac model or the multi-time approach for relativistic BM. Finally, Section \ref{conc} concludes the essay.

\section{Main Formulations of the Pilot-Wave Theory}
\label{PWT}

\subsection{Louis de Broglie's Original Proposal}
\label{db}
The pilot-wave theory originated from the fundamental ideas contained in de Broglie's PhD thesis and his subsequent essays (\cite{Broglie:1924, Broglie:1925, Broglie:1927}). Between 1923 and 1925, de Broglie developed the revolutionary hypothesis that any physical object exhibits both corpuscular and undulatory properties. Thus, \emph{matter} can be thought of not only in corpuscular terms but also as having a wave-like essence.\footnote{De Broglie was awarded the Nobel Prize in physics in 1929 for his hypothesis about the wave nature of electrons.} 
Indeed, to each particle of mass $m$ and momentum $p$, he associated a wave of wavelength $\lambda = h/p$ and a frequency $\nu=E/h$. Consequently, the wave-particle duality became a general feature of nature, generalizing Einstein's wave-particle description of light.
 Notably, with this new proposal, the French physicist was able to explain and predict a wide range of experimental phenomena, such as the interference of single photons, the diffraction and interference of electrons---as verified experimentally in the mid-1920s by C.J.\ Davisson, L.\ Germer, G.\ Thomson, and A.\ Reid---or the Bohr-Sommerfeld quantization conditions of atomic energy levels.

Moreover, de Broglie proposed a new dynamics according to which particles' velocities were determined by the gradient of the phase of such an associated wave. 
In his theory, Newton's principle of inertia was rejected in favor of a new one that ``unified Maupertius' variational principle for mechanics with Fermat's variational principle for optics'', as underlined by \cite{Bacciagaluppi:2009}, p.\ 37, aiming at synthesizing the physics of particles with that of waves (cf.\ \cite{Broglie:1925}).
In particular, according to this proposal, particles' velocities are determined by the gradient of the phase of the accompanying wave, contrary to the classical Newtonian theory in which forces determine accelerations. Another interesting aspect of de Broglie's theory is that quantum particles have definite positions and follow deterministic paths in space. This fact marks an important difference with respect to other research programs developed in the 1920s, for matrix and wave mechanics both dismissed the particle concept in their ontologies---and consequently the notion of trajectory. 

However, we should underline that the PWT was an approximation of a more general project, namely de Broglie's ``double solution'' programme, whose ``core idea is that there should be two synchronous, coupled solutions of the wave equation'' (\cite{Colin:2017}, p.\ 26). The double solution theory introduces a $\psi$-wave, of which only its phase carries physical significance, and a $u$-wave, which is a solution representing a ``singularity'' that corresponds to a particle which follows the flow lines of Schr\"odinger's wave equation. 
At the beginning of the double solution program, de Broglie assigned to each $u$-wave the position of a corpuscle, and postulated that the particles' velocities were guided by the phase of the wave function $\psi$. In this way, he created the basis of the pilot-wave theory that was presented at the fifth Solvay conference in 1927. 

On that occasion, after having explained the essential elements of his new ideas concerning the corpuscular and undulatory nature of quantum objects, de Broglie illustrated a new dynamics for quantum objects and a new interpretation for the wave function. The French scientist described a spinless one-particle system with a wave function $\psi(x, t)$ whose dynamical evolution in the non-relativistic case is governed by the Schr\"odinger equation\footnote{For technical details on the original de Broglie's memoir, cf.\ Bacciagaluppi and Valentini (2009), pp.\ 374-407.}:
\begin{align}
\label{SEq}
i\hbar\frac{\partial }{\partial t}\psi(x,t)=\bigg(-\frac{\hbar^2}{2m}\nabla + V\bigg)\psi(x,t). 
\end{align}
\vspace{1mm}

\noindent As already mentioned above, he hypothesized that such a particle always has a definite position in space and moves along a deterministic trajectory guided by the $\psi$ wave. The following law of motion determines the particle's path: 
\begin{align}
\label{dB}
\frac{d x}{dt} = \frac{1}{m}\nabla S, 
\end{align}
\vspace{1mm}

\noindent where $S$ is the phase of the guiding wave. 

According to de Broglie's theory, the wave function has two roles: it guides the motion of the corpuscles---hence the name `pilot' wave---and it determines the probability, in agreement with Born's rule, that a given particle occupies a certain location in space at an arbitrary time. As we will see, such a characteristic of $\psi$ will be maintained in future pilot-wave theories.

Bacciagaluppi and Valentini's detailed analysis of the subsequent general discussion that followed de Broglie's talk underlines that he was not fully able to reply to Kramers' objection involving the recoil of a photon hitting a mirror. To answer it, the pilot-wave dynamics should have taken into account not only the microscopic system, the photon, but also the mirror. Similarly, Pauli's objection concerning elastic scattering was not completely responded to by de Broglie. As is well known, by 1930, the French scientist abandoned the pilot-wave approach together with his double solution theory.\footnote{Various historical accounts explain de Broglie's abandonment of the PWT. Some authors attribute a crucial role to Pauli's criticisms (cf.\ \cite{Jammer:1974aa}, \cite{Freire:2019}, and \cite{Talbot:2017}), while \cite{Bacciagaluppi:2009} argues that they were not decisive, as de Broglie continued to develop his program until 1930. Later, especially after Bohm's work in 1952 and its recognition in France, de Broglie revived his double solution theory, attempting to introduce nonlinear modifications to the Schr\"odinger equation.} 

\subsection{David Bohm's rediscovery of Pilot-Wave Theory}
\label{Bohm}

After publishing the well-received textbook {\em Quantum Theory} in 1951 (\cite{Bohm:1951}), David Bohm became dissatisfied with the standard interpretation of QM and, in the same year, developed his own version of PWT (\cite{Bohm:1952aa, Bohm:1952ab}).\footnote{In 1951, Bohm was unaware of de Broglie's work. Pauli pointed it out to him in private correspondence. Bohm eventually gave due credit to the French physicist for introducing the idea of pilot-wave theory, but also said that he gave up the theory ``too soon''. On this episode, cf.\ \cite{Freire:2019}, pp.\ 71-73.} 
Bohm's primary aim was to show that another interpretation of the quantum formalism could be given, despite the extraordinary empirical adequacy and predictive power of QM (\cite{Bohm:1952aa}, p.\ 168). In particular, he demonstrated that a causal and ontologically unambiguous description of quantum objects and processes was obtainable, contrary to the a-causal picture of the standard theory. 

In his essays, Bohm achieved several results that still constitute the core of any PWT. One of his major contributions is providing a theory of quantum measurements that can explain dynamically and causally how experimental results are actualized and how they emerge from the motion of individual quantum particles guided by a wave field, filling the explanatory gap left by standard quantum theory as well as by de Broglie's original proposal (cf.\ \cite{Bohm:1952ab} Section 2).\footnote{\cite{Bacciagaluppi:2009} illustrated also how to answer Kramers' objection using Bohm's theory of measurement, cf.\ Chapter 10.4, pp.\ 246--247. Moreover, Bohm answered Pauli's objection both publicly, \cite{Bohm:1952aa}, Section 7, and in Appendix B of \cite{Bohm:1952ab}, as well as privately, as discussed in detail in \cite{Oldofredi:2023}.}
Moreover, he clarified that observables should not be interpreted as properties of quantum systems, but rather as describing an interaction between the object system under consideration and the experimental set-up involved (cf.\ {\em ibid}. Section 4 and Section \ref{measure} below).

According to Bohm's approach, quantum systems are described in terms of particles whose position is defined at all times, dynamically guided by wave functions--- considered physically real fields---whose time evolution is governed by the Schr\"odinger equation.\footnote{The wave-particle duality manifests itself in a new fashion in Bohm's account, for he introduces a dualistic ontology of particles \emph{and} wave fields. Those physical situations in which we observe undulatory features of the quantum particles, as in the famous double-slit experiment, are explained via the action of the $\psi$-field in space.} Thus, in this theory, it is not the case that $\psi$ alone provides the complete description of a certain system.

Moreover, Bohm's theory boils down to a re-interpretation of the Schr\"odinger equation, from which the velocity of quantum particles can be derived, and thereby the notion of trajectory can be retained. Consequently, he showed that (i) his interpretation does not entail {\em any} modification to the standard quantum formalism, (ii) that the very notion of trajectory is already embedded in the mathematical framework of QM\footnote{Cf.\ Section \ref{measure} for details.}, and (iii) that the notion of causality can be reintroduced in the realm of quantum physics.

To define his theory, Bohm proceeded as follows. First, the wave function is expressed in polar form $\psi=Re^{(iS/\hbar)}$---where $R$ and $S$ represent two coupled real functions corresponding to the amplitude and the phase of the wave, respectively. Second, posing $P(x)=R^2(x)$, where $P(x)$ represents the probability density, one obtains the following equations for $R$ and $S$:

\begin{align}
\label{density}
\frac{\partial P}{\partial t} + \nabla\cdot\left(P\frac{\nabla S}{m}\right)=0, 
\end{align}

\begin{align}
\label{QHJ}
\frac{\partial S}{\partial t} + \frac{(\nabla S)^2}{2m}+V(x)-\frac{\hbar^2}{4m}\Bigg[\frac{\nabla^2 P}{P}-\frac{1}{2}\frac{(\nabla P)^2}{P^2}\Bigg]=0. 
\end{align}

\noindent The former is the quantum continuity equation for the probability density, whereas the latter is the \emph{quantum Hamilton-Jacobi} equation describing the motion of a particle (or a configuration of particles) with kinetic energy $(\nabla S)^2/2m$ and subjected to the influence of both a classical and a new \emph{quantum} potential:

\begin{align}
\label{Q}
Q=\frac{-\hbar^2}{4m}\Bigg[\frac{\nabla^2 P}{P}-\frac{1}{2}\frac{(\nabla P)^2}{P^2}\Bigg]=\frac{-\hbar^2}{2m}\frac{\nabla^2R}{R}. 
\end{align}

\noindent This latter element is the distinctive aspect of Bohm's formulation of PWT, absent in de Broglie's original formulation\footnote{Another remarkable difference is that de Broglie's theory implements a first-order dynamics, in which velocities are the fundamental quantities, whereas Bohm formulates a second-order theory, where accelerations (as e.g., the quantum potential) and forces are primary. Indeed, Bohm's theory can be casted in Newtonian fashion, making explicit the actions of the classical and the quantum potentials on the quantum particles, cf.\ \cite{Bohm:1952aa}, pp.\ 170-171.}, and it plays a crucial role in explaining the new features arising in the quantum domain---as the interference pattern of the double-slit experiment, the EPRB spin correlations, \emph{etc.}.

The quantum potential indeed possesses salient attributes not present in the classical case: (i) it depends only on the form of the wave function, meaning that it is insensitive to the intensity of $\psi$, and thereby even a weak quantum field can strongly affect the particles, (ii) its action does not decrease with the distance separating the corpuscles, generating a non-local dynamics (in agreement with Bell's theorem), and (iii) it contains information about the whole experimental set-up.

In his 1952 papers, Bohm defined the particles' velocity as 

\begin{align}
\label{Bv}
v=\nabla S/m
\end{align}

\noindent to the effect that his proposal is metaphysically unambiguous, for the Bohmian corpuscles always have a precise localization in three-dimensional space and a well-defined velocity independently of any observation or measurement. 
Referring to this, Bohm interestingly stated that in his theory the uncertainty principle becomes just ``an effective practical limitation on the possible precision of measurements'' (\cite{Bohm:1952aa}, p.\ 171). 
Hence, it should not be interpreted as an irreducible impossibility to conceive the concepts of position and momentum as simultaneously defined quantities, as Heisenberg instead famously claimed. It is relevant to underline that in virtue of the identity $v=\nabla S/m$ one can rewrite the quantum continuity equation for the probability density as $\partial P/\partial t + \nabla\cdot\left(Pv\right)=0$. where $Pv$ is interpreted as the mean current of the particles in a given configuration. This equation is particularly important since the Born distribution holds as a consequence of it.\footnote{\cite{Bohm:1952ab} Section 9 explains why von Neumann's theorem does not refute the causal interpretation. In a nutshell, von Neumann's result shows that there cannot be dispersion-free states (cf.\ footnote 1). However, Bohm underlines that in his theory, quantum observables are ``not properties belonging to the observed system alone, but instead potentialities whose precise development depends just as much on the observing apparatus as on the observed system''. We will see this feature of Bohm's theory in detail in Section \ref{measure}. Moreover, in private correspondence, Bohm wrote that von Neumann himself recognized the logical consistency of his theory. Cf.\ Bohm's letter to Pauli in \cite{Meyenn:1996}, Letter 1290, p.\ 392 and Bohm's letter to Melba Phillips in early 1952 in \cite{Talbot:2017}, p.\ 147.}

Before closing this section, it is worth noting that in Bohm's view, the causal interpretation---as he later called it---was only the first step to reach a satisfactory description of the microscopic regimes of our universe, not a proposal for a fundamental ontology of the quantum (contrary to some contemporary Bohmian approaches, cf.\  \cite{Esfeld:2017}). By looking at Bohm's scientific and philosophical production, in fact, he constantly revised, modified, and put into question his own perspectives on the meaning of quantum theory, providing various conceptions of his own hidden variable model.\footnote{For a comprehensive account of Bohm's approach to quantum foundations, the reader is referred to \cite{Freire:2019}.} Therefore, his approach should be viewed as a preliminary step towards a better understanding of the quantum realm. 

Referring to this, Bohm did not believe that a certain ontology---in particular, a corpuscular ontology---should necessarily be maintained at all energy/length scales. Indeed, in Appendix A to his 1952 papers, he provided a guidance equation for {\em field} coordinates describing the electromagnetic field (cf.\ Section \ref{field}).
On the other hand, already in his correspondence with Pauli in late 1951 and in Section 9 of \cite{Bohm:1952aa}, he made non-trivial statements about the possibility of modifying his own interpretation, showing a willingness to develop and generalize it beyond the scales in which non-relativistic QM is applied.\footnote{Bohm claimed in several places that his own approach, exactly as QM, has limited validity. See \cite{Bohm:1953aa} and \cite{Bohm:1957}, as well as his private correspondence published in \cite{Talbot:2017}. This is a characteristic trait of Bohm's conceptions of physical theories, which, in his opinion, always had a specific, restricted domain of application. More details on Bohm's philosophy of science can be found in \cite{Oldofredi:2023}.} Moreover, Bohm was convinced that his approach would have entailed new physics at very short length scales, where he also conceived the possibility of formulating a {\em sub-quantum} theory implementing a non-linear dynamics. Interestingly, he speculated that such a deeper, more fundamental nonlinear law applied at very short distances would allow for a precise description of matter beyond the threshold imposed by current quantum theory, envisioning experiments with unlimited predictions able to test the predictions of the pilot-wave theory, making it \emph{falsifiable} and thereby testable against the predictions of the standard formulation of QM. Although such speculations did not concretize into new results, Bohm's hope to make his theory falsifiable or testable remains open in today's proponents of the pilot-wave theory (for more details, cf.\ Sections \ref{philo} and \ref{Spin}). 

\subsection{Bohmian Mechanics}
\label{BM}

Bohmian Mechanics is a theoretical framework developed since the early 1990s by Detlef Dürr, Sheldon Goldstein, and Nino Zanghì (\cite{Durr:2013aa}). 
Building on John Stuart Bell's reformulation of de Broglie-Bohm theory with a first-order dynamics (cf.\ \cite{Bell:1982}, p.\ 992, and footnote 16), these physicists provided a deterministic quantum theory of particles that move in three-dimensional physical space and follow continuous trajectories. 

According to this theory, physical systems are described by a pair $(\psi, Q)$, where the former is the quantum mechanical wave function defined in configuration space $\mathbb{R}^{3N}$ and the latter represents a specific $N$-particle configuration with positions $(Q_1,\dots, Q_N)$ in physical space. As customary in PWT, two dynamical laws are given: the usual Schr\"odinger equation prescribing the evolution of the wave function $\psi=\psi(q_1,\dots, q_N, t)$
\begin{equation}
\label{SE}
i\hbar\frac{\partial{\psi}}{\partial{t}}=-\sum^{N}_{k=1}\frac{\hbar^2}{2m_k}\Delta_k\psi +V\psi 
\end{equation}

\noindent{and}, on the other hand, the guiding equation governing the particles' motion:
\begin{equation}
\label{guide}
\frac{dQ_k}{dt}=\frac{\hbar}{m_k}\mathrm{Im}\frac{\nabla_k\psi}{\psi}(Q_1,\dots, Q_N)=v_k^{\psi}(Q_1,\dots, Q_N).
\end{equation}
 
\noindent The vector velocity field on the r.h.s.\ of \eqref{guide} depends on the wave function, whose role is to guide the motion of the particles. This law provides the evolution of the $k$th particle at time $t$ depending on the positions of all the other particles at that time. It appears immediately evident how BM differs significantly with respect to Bohm's 1952 theory, for in the former theory the quantum potential is eliminated in favor of a simpler mathematical structure. 

The empirical equivalence between BM and standard QM is achieved via \emph{equivariance}: if we assume that at any arbitrary initial time $t_0$ the particle configuration is distributed according to $|\psi_{t_0}|^2$, then it will be so distributed for any later time $t>t_0$, preserving the Born's distribution (see \cite{Durr:2013aa}, Chapter 2, Sec. 7 for the mathematical justification of this statement). More precisely, based on the universal wave function $\Psi$, a unique equivariant typicality measure can be defined in terms of the $|\Psi|^2$-density. Given that typicality measure, it can then be shown that for the overwhelming majority of initial conditions, the distribution of particle configurations in an ensemble of sub-systems of the universe that admit of a wave function $\psi$ of their own (known as {\em effective wave function}) is a $|\psi|^2$--distribution. A universe in which this distribution of the particles in sub-configurations obtains is considered to be in quantum equilibrium.

As we will see in more detail in Section \ref{wf}, another relevant metaphysical difference between BM and Bohm's 1952 theory consists in the interpretation of the wave function: whereas Bohm interpreted it as a real physical field, for D\"urr, Goldstein, and Zanghì, $\psi$ is not a physical object, but rather a nomological entity. This becomes clear if we consider the methodological framework that motivates BM, namely the Primitive Ontology (PO) approach, a normative methodology for theory construction. 

Following Bell's theory of local beables (\cite{Bell:1975aa}), the PO programme requires that the mathematical structure of a theory $T$ be divided into two subcategories. First, the ``primitive'' variables representing matter, and referring directly to real objects precisely localized and moving in 3-dimensional space (or in space-time), which is generally considered a real substance as well.
These constitute the primitive ontology of the theory, i.e., the fundamental entities postulated by $T$ which constitute the building blocks of macroscopic reality. Second, $T$ contains mathematical structures that are responsible for the dynamical evolution of the PO, without representing material objects---the non-primitive (or non-physical) variables. The wave function belongs in the second category.

This fact entails that, contrary to Bohm's theory, BM implements a monistic ontology with respect to material objects. Moreover, while Bohm did not believe in a strong form of reductionism (as explicitly stated in his monograph \cite{Bohm:1957}), the PO approach endorses a reductionist view according to which everything that exists is reduced to the motion in space of the PO. In the case of BM, everything is explained in terms of particles' positions and trajectories.\footnote{A detailed examination of the differences between the metaphysical projects of Bohm, Bell, and the PO theorists is given in \cite{Oldofredi:2022}.} 

\subsection{Non-standard formulations}

In addition to the main formulations of PWT introduced above, there are also various non-standard versions of BM, which have been proposed to address particular metaphysical or technical issues affecting the standard account of BM. In what follows, a few relevant examples are briefly discussed.

\begin{itemize}
\item [] {\bf Identity-Based Bohmian Mechanics:} This theory proposes a radical metaphysical hypothesis concerning the nature of Bohmian corpuscles. According to this framework, there are no different particle species but just one family of ``naked'' particles. Thus, there is no intrinsic metaphysical difference between an electron, a positron, or a muon; they are instances of the very same family of identical particles. In this view, the Bohmian particles do not possess inherent properties such as mass, charge, or spin. This is reflected in the permutation-invariant equations governing their motions, with the effect that one cannot distinguish them {\em a priori}, but only by looking at their actual paths in space. In fact, these objects happen to move differently following diverse trajectories; such dynamical behavior is what allows us to distinguish them---cf.\ \cite{Goldstein:2005a, Goldstein:2005b} for technical details. In \emph{Identity-based Bohmian mechanics}, ``intrinsic'' attributes are then included in the wave function as dynamical parameters of the theory. Based on this framework, Esfeld and Deckert developed a relationalist account of BM (\cite{Esfeld:2017}).

\item [] {\bf Retrocausal BM:} In a series of papers, Roderick I.\ Sutherland provided a retrocausal account of BM. In this framework, future states $\psi_f$, i.e., final boundary conditions (cf.\ \cite{Friederich:2019}) are added to the usual wave function $\psi_i$, where the index $i$ refers to arbitrary initial conditions. Sutherland derives a guiding equation where such final states may influence particles' trajectories in the past. The main goal of the author is twofold: on the one hand, to make the Bohmian theory local, and on the other hand, to preserve Lorentz invariance. For details, cf.\,\cite{Sutherland:2008, Sutherland:2015, Sutherland:2017}; a critical analysis of this model can be found in \cite{Berkovitz:2008}.

\item [] {\bf Non-Standard Analysis BM:} In a recent essay \cite{Barrett:2023}, BM is formulated with nonstandard analysis to describe a broader set of physical systems with respect to the standard version. Moreover, nonstandard BM can explain Earman's space invaders (a hypothetical situation in which a particle is accelerated in a way to go to infinity in a finite time), whereas the classical theory cannot. Additionally, considering eq.\ \eqref{guide}, it is well-known that at those points where $\psi$ vanishes (i.e., nodes), the denominator is zero; thereby, the particles' velocities diverge to infinity or become indeterminate. This fact entails interruptions in the particles' trajectories. To tame this issue, the standard theory assumes, e.g., that the set of nodes has zero Lebesgue measure. However, within Barrett and Goldbring's framework, one may dismiss such an assumption, for the value of $\psi$ at a nodal point may not be zero, but can instantiate a non-zero infinitesimal, that is, a number which has an infinitely small value, but not zero. Therefore, particles' trajectories are not interrupted and can pass through singularities.
\end{itemize}

\section{Philosophical Intermezzo}
\label{philo}
\subsection{Born's Rule and the Interpretation of Probability}

Although the main formulations of PWT illustrated in the previous section show remarkable differences in the dynamics of quantum particles or in the meaning of $\psi$ (as we will see later in this section), these frameworks share a common interpretation of quantum probability. In such theories, quantum probabilities are mere manifestations of ignorance about the exact particles' positions on the one hand, and of the empirical fact that one cannot manipulate them operationally on the other. Thus, the maximum knowledge at our disposal is given by the Born distribution. In pilot-wave theory, then, probabilities do not refer to an inherent indeterminacy of quantum objects, but rather possess an epistemic character. Since we can neither know nor manipulate the initial positions of the particles in experimental situations, measurement outcomes will appear random.

Nonetheless, despite such a common reading of quantum probability, many authors had different views concerning the status of Born's rule and how to derive it. In what follows, I introduce the three major positions in the literature. However, as the reader will see, this question has not yet achieved a definitive, satisfactory answer.\footnote{A recent, interesting criticism of BM and its interpretation of probability is contained in \cite{Ruetsche:2023}.} 

\subsubsection{Bohm's Attempts to Derive Born's Rule}

After the publication of his 1952 papers, Bohm continued to work on his causal interpretation and sought to show how Born's rule dynamically emerges from the random collisions of (sub)quantum particles, for he viewed QM as a statistical theory emerging from a deeper, subquantum regime (\cite{Bohm:1952aa}, Section 9). Indeed, \cite{Bohm:1953b} argues that if the particles' velocities satisfy the condition $v= \nabla S/m$ and the wave function evolves according to \eqref{SE}, then {\em any arbitrary} probability distribution will eventually approach $|\psi|^2$ due to a large number of random collisions. Here he claims (p.\ 274) that such a result is ``the direct analogue of Boltzmann's $H$-theorem in classical statistical mechanics''. Therefore, Born's distribution should no longer be postulated as an additional axiom, but rather it would be a consequence of its ontology and dynamics. This argument, however, has been strongly criticized since ``no general argument for relaxation was provided'' (\cite{Valentini:2025}, p.\ 72).

Similarly, \cite{Bohm:1954aa} aimed at achieving the same result using a hydrodynamic model---originally proposed by Madelung in 1926. The authors considered a fluid where $|\Psi|^2$ represents its density, and $\nabla S/m$ refers to the local stream velocity. 
However, the equations \eqref{density} and \eqref{QHJ} assumed a new meaning with respect to Bohm's 1952 theory, now representing respectively the conservation of the fluid and the law determining the velocity potential, which includes in this case both classical and quantum forces. An additional assumption was also made, i.e.\ that the fluid fluctuates irregularly to resemble turbulent motion, where the actual density $\rho$ and the actual velocity $v$ randomly fluctuate around $|\Psi|^2$ and $\nabla S/m$, which are taken as average quantities. 
Such irregular motions and fluctuations of the fluid's elements produce a ``mixing'' of the fluid itself---in which each ``particle'', or better, body-like inhomogeneity, follows a very irregular path. The authors then showed that a statistical ensemble of such systems with an arbitrary probability density $P(x)$ decays into $P=|\Psi|^2$, concluding that the irregular fluctuations of the fluid provide a physical explanation for the statistical distributions of QM.

Although interesting, many authors consider this latter proposal {\em ad hoc} since it relies on many weak assumptions concerning subquantum stochastic motions, and artificial, for quantum relaxation occurs also in de Broglie's deterministic dynamics, without the need to introduce a stochastic framework, cf.\ \cite{Valentini:2025} and \cite{Drezet:2021}.

\subsubsection{Quantum Relaxation and Non-Equilibrium}
\label{Val}

Along the lines of Bohm's attempts to justify Born's rule, Anthony Valentini proved a sub-quantum $H$-theorem affirming that $|\psi|^2$ emerges on a coarse-grained level from a process of quantum relaxation starting with an initial state of non-equilibrium, in which $\rho\neq |\psi|^2$ (\cite{Valentini:1991a, Valentini:1991b}). On Valentini's view, this can be compared to thermal equilibrium in classical physics, which ``can be understood as arising from a process of thermal relaxation'' (\cite{Valentini:2025}, p.\ 70). Thus, he aims at proving a dynamical explanation for the $|\psi|^2$-distribution; however, contrary to Bohm's strategy, he assumed the validity of the deterministic pilot-wave dynamics (cf.\ also \cite{Valentini:2005aa, Valentini:2025}). Moreover, Valentini often claims that quantum relaxation can be viewed as a mixing of two fluids with densities respectively $\rho$ and $|\psi|^2$ ``in configuration space, which obey the same continuity equation and are ‘stirred’ by the same velocity field $\dot{q}$, eventually becoming indistinguishable on a coarse-grained level'' (\cite{Valentini:2025}, p.\ 71).

This view, as \cite{Norsen:2018} underlines, ``can be understood as a justification of the quantum equilibrium hypothesis,   In simple terms, the idea being that even if, say, back at the big bang, particle positions were {\em not} $P_B$-distributed [Born-distributed], they would inevitably become $P_B$-distributed (at least in a good-enough-for-all-practical-purposes, FAPP, coarse-grained sense) during the subsequent dynamical evolution of the universe, thereby justifying the application of the QEH to the ``initial'' conditions relevant to contemporary experimental investigations''.\footnote{For a detailed presentation and critical assessment of this approach, see \cite{Norsen:2018}, Sections 3 and 4. Other objections against Valentini's perspective are given in \cite{Callender:1997}.}

In a series of papers, Valentini not only claimed that one can derive Born's rule without including it among the postulates of PWT---thereby simplifying its structure---but also that his perspective would lead to new physics, making this framework experimentally falsifiable. For technical details, cf.\ \cite{Valentini:2025} Sections 2.2.2-2.2.4 and 2.5.1.

Despite the undeniable interest of Valentini's view, many philosophers and physicists severely criticized his approach. For instance, \cite{Drezet:2021} underlines that the mixing and coarse-graining techniques employed in the derivation of Valentini's sub-quantum $H$-theorem cannot be used to derive relaxation. In addition, ``one cannot show that the entropy is a monotonously growing function ultimately reaching quantum equilibrium'', a point also made in \cite{Callender:2007aa}.

\subsubsection{Quantum Equilibrium}

Contrary to Bohm and Valentini's proposals, \cite{Durr:1992aa} justifies the so-called {\em quantum equilibrium hypothesis}, a postulate introduced in BM to guarantee the theory's empirical adequacy and its empirical indistinguishability from standard QM. These authors show that the Bohmian dynamics is {\em equivariant}, that is, if at an arbitrary system in configuration space at some initial time a particle configuration is Born-distributed, it will be so distributed for any later time (cf.\ \cite{Bricmont:2016aa}, Chapter 5, \cite{Goldstein:2025}). 

Moreover, if it is assumed that the actual universe is a typical Bohmian universe, i.e., one that is in quantum equilibrium, Born's rule can be derived for the calculation of measurement outcome statistics on {\em sub-systems} of the universe. To do so, it must be postulated that $|\Psi|^2$ holds, providing a typicality measure for the universe, where $\Psi$ is the universal wave function. This axiom, in turn, justifies the application of the $|\psi|^2$ rule for the calculation of the probabilities of measurement outcomes on particular sub-systems of the universe, with $\psi$ being the effective wave function of the particular systems under consideration. 

Although supporters of PWT have largely endorsed this view, it can be critically assessed in various ways. On the one hand, it should be emphasized that Born's rule is neither explained nor derived {\em for the universe}, for D\"urr, Goldstein, and Zangh\`i postulated that a typical Bohmian universe is $|\Psi|^2$-distributed. Therefore, one must say that \cite{Durr:1992aa} justifies the use of Born's rule for {\em subsystems} of the universe, rather than having provided a full justification for the Born rule per se, and thus having derived it from other principles of the theory. On the other hand, Valentini claimed in several places that such a view does not entail new physics, contrary to his approach. Moreover, he underlines that whereas laws of nature are stable, fixed features of our universe, the same is not true for initial conditions, which are merely contingent and should be determined empirically, not postulated a priori. In his words, ``[w]hether our universe began in equilibrium
or in nonequilibrium is an empirical question to be determined by cosmological
observation'' (\cite{Valentini:2025}, p.\ 72).\footnote{Further criticisms of this view can be found, e.g., in \cite{Allori:2021}, \cite{Callender:1997},  \cite{Callender:2007aa}.}

\subsection{Major Perspectives about the Wave Function}
\label{wf}

\subsubsection{Bohm on the Meaning of $\psi$}

As we briefly mentioned in Section \ref{Bohm}, Bohm provided a realist understanding of $\psi$ in his 1952 work on pilot-wave theory: rather than considering it as a formal device storing information about probabilities of measurement outcomes, it was interpreted as a physically real field, exactly as the electromagnetic field. In fact, Bohm proposed compelling analogies between these two entities, underlining, e.g., that the $\psi$-field exerts a force on the particle similar to the way in which an electromagnetic field exerts a force on a specific charge.\footnote{This analogy is maintained in later volumes as e.g., \cite{Bohm:1957, Bohm:1993aa}.}
Additionally, he claimed that\footnote{Bohm was obviously aware of the important differences between the two---as e.g.\ the lack of back reaction of the particles on the $\psi$-field, or the absence of sources for the wave function.}

\begin{quote}
just as the electromagnetic field obeys Maxwell's equations, the $\psi$-field obeys Schr\"odinger's equation. In both cases, a complete specification of the fields at a given instant over every point in space determines the values of the fields for all times. In both cases, once we know the field functions, we can calculate force on a particle, so that, if we also know the initial position and momentum of the particle, we can calculate its entire trajectory (\cite{Bohm:1952aa}, p.\ 170). 
\end{quote}

However, Bohm became increasingly dissatisfied with this interpretation of the wave function because, in a many-body system, $\psi$ is defined in a multi-dimensional configuration space; therefore, it can no longer be considered a field exerting a mechanical force on the particles in three-dimensional space. 

To tame this problem while maintaining a realist reading of $\psi$, in the 1980s Bohm and Hiley developed a new interpretation of the wave function in terms of `information'. They maintained that $\psi$ is still interpreted as a real physical field even though it does not ``push'' the particles in space; rather, it carries information about their motion (cf.\ \cite{Bohm:1990}, \cite{Bohm:1982, Bohm:1985, Bohm:1993aa}).\footnote{Recently, \cite{Pylkkanen:2025} underlines that already in the 1960s Bohm was dissatisfied with another aspect of his realist view of $\psi$, for it gives rise to the quantum potential whose mathematical form seems arbitrary. However, Pylkk\"anen says, re-examining his 1952 work during the 1970s and 1980s with Hiley, Bohm ``discovered a way to make sense of the strange mathematical form of the quantum potential. The fact that the quantum potential depends on the second spatial derivative of the quantum field means that the quantum potential only depends on the {\em form} of the quantum field. Bohm developed this into the idea that the quantum field is literally in-forming (i.e., putting into form) the movement of the particle, rather than pushing and pulling it mechanically. [...] Thus in Bohm's view not only does the concept of active information alleviate the feeling that the mathematical form of the quantum potential is ``strange and arbitrary '', it also provides a way to make sense of the multidimensional nature of the many-body wave function''.} Bohm and Hiley wrote that

\begin{quote}
[t]he fact that the wave function is in configuration space clearly prevents us from regarding the quantum field as one that carries energy and momentum
that was simply transferred to the particles with which it interacted (thus effectively pushing or pulling mechanically on the latter). This is a further factor in
addition to the form dependence of the activity of the field, which leads us to consider the interpretation of this field as active information (\cite{Bohm:1993aa}, p.\ 61).
\end{quote}

\noindent The notion of information here employed is not related to any technical definition of it, as e.g.\ Shannon's theory. They interpreted the word ``information'' in the sense of ``giving form to something'', following its Latin root. 
In this new perspective, the wave function informs---i.e.\ gives form to---the particles' trajectories. Contrary to Bohm's previous works, the wave field does not act directly on the particles in space, but rather the form of $\psi$ (in particular the second spatial derivative of the amplitude $R$ appearing in the quantum potential) prescribes the dynamical evolution of the corpuscles. To better understand this point, Bohm offers the following analogy:

\begin{quote}
consider a ship on automatic pilot guided by radar waves. The ship is not pushed and pulled mechanically by these waves. Rather, the form of the waves is picked up, and with the aid of the whole system, this gives a corresponding shape and form to the movement of the ship under its own power (\cite{Bohm:1990}, p.\ 279).
\end{quote}
 
\noindent In this example, the radio waves carry information about the environment and guide the ship through it. Similarly, the $\psi$-wave---and in particular the quantum potential---contains information about specific experimental environments and guides the motion of the particles through it. 

Notably, this concept of information is strictly related to the notion of {\em wholeness}. As we already mentioned in Section \ref{Bohm}, the action of the quantum potential does not in general vanish with distance: namely, it is independent of the intensity of the quantum field, but depends on its form. It is its influence on the particles---those composing the observed system as well as the measuring apparatus---that determines and explains observable measurement outcomes. Hence, Bohm claimed, these features of $Q$ fact imply a holistic relation between the observed system and its environment.\footnote{This reading of $\psi$ holds that the information of the wave function is potentially active in every spatial region, but it is active only where particles are located. Empty wave packets are now defined as {\em potentially active information}, since the particles may interfere again with them in their future paths, and thereby they can become active and effectively guide their motion.} Additionally, to fully understand the notion of multidimensional information, and thereby the notion of multidimensional reality, entailed by Bohm's ontological commitment towards the wave function---and introduced in \cite{Bohm:1980}---one has to embrace the notion of {\em implicate order}, as correctly emphasized by \cite{Pylkkanen:2025}. In a nutshell, Bohm's notion of implicate order should be understood as the most basic, sub-quantum layer of reality, from which our sensorial, phenomenal experience emerges (the explicate order). This concept expresses the idea that the whole universe in its entirety is to be found---or in Bohm's jargon, is ``enfolded''---in everything, and that each object is enfolded in the whole. In this framework, quantum wholeness assumes a central role through the quantum potential. Since our explicated, experienced 3D space does not have enough structure to track the holistic nature of our quantum universe, a multi-dimensional, implicate notion of reality is hypothesized. 

Since the ideas of the late Bohm, in particular the theory of the implicate/explicate order, received harsh criticisms and did not find support among the contemporary supporters of PWT, his views became secondary and neglected.\footnote{For recent studies on Bohm's late philosophical positions, cf.\ \cite{Knuuttila:2025}, \cite{Pylkkanen:2025}, \cite{vanStrien:2020}.}

\subsubsection{The Nomological View}
\label{nom}

The \emph{nomological} interpretation of the universal wave function $\Psi$ is proposed by \cite{Goldstein:2013}. According to this view, the wave function of the universe is not a physical item existing in spacetime that causally guides the particles through it, nor is it a multidimensional field. Rather, it is an abstract entity that governs their motion, \emph{generating} the vector velocity field of the corpuscles. To clarify what this means, the authors claim that $\Psi$ plays in BM a role analogous to the Hamiltonian function in classical mechanics. The latter generates a vector field in classical phase space via Hamilton's equations; similarly, the wave function generates a vector field in configuration space via the guiding equation of the Bohmian theory. In the same vein, wave functions describing subsystems of the universe---for instance, those representing experimental situations---are defined as \emph{quasi-nomological}, inheriting the nomic character from the universal wave function.\footnote{A critical analysis of this latter notion is given in \cite{Oldofredi:2021}. Objections to the nomological view are given, e.g., in \cite{Belot:2012aa}, \cite{Allori:2021}.}
Notably, although $\Psi$ is not a physical object, it is part of the ontology of BM, which includes both primitive and non-primitive variables, as already mentioned in Section \ref{BM}; thus, some ontological weight is assigned to the wave function.
Referring to this, the authors are realist about laws of nature; thus, even though $\Psi$ is not a material substance, it is considered a real abstract entity in the Bohmian ontology.

Along the lines of the nomological view, \cite{Esfeld:2017} formulate a relationalist ontology of propertyless, finite, and permanent matter points uniquely individuated by their distance relations---the only property of these points being their position.  The change of these relations, which is taken as a primitive fact, constitutes the dynamical aspect of the proposal. Thereby, BM becomes a theory about the evolution of configurations of matter points and their changes in distance relations. 
Conforming to this minimalist ontology, the wave function is not a physical object, but a part of the dynamical structure of the Bohmian theory. 

However, Esfeld and Deckert support a different metaphysics about $\Psi$ (and $\psi$) with respect to Goldstein and Zangh\`i, for they endorse Humeanism about laws of nature. From their perspective, then, the world is just a mosaic of particular, local facts about matter points and their distance relations. Moreover, in the Humean view, there is nothing in the world that constrains the dynamics of these points: a given temporal evolution just contingently happens to occur. The patterns and regularities exhibited by the dynamical change of matter points are conceptualized via the Humean best system account, according to which physical laws are the axioms of ``the system that achieves the best balance between being simple and being informative in describing the evolution of the configuration of matter'' (\cite{Esfeld:2017}, p.\ 45). For instance, in Esfeld and Deckert's opinion, the Humean mosaic is given by the distribution of matter points whose progressive change shows regularities described by the laws of Bohmian mechanics, which are assumed to be the best description of such a mosaic at the non-relativistic level. In this scheme, the wave function is an important representational tool in the laws of the theory that efficiently describes the temporal enfolding of the matter points. Thus, in their view $\psi$ is a ``central dynamical parameter figuring in the law of motion for the particles'' (\emph{ibid}., pp.\ 51--52).

\subsubsection{Other Proposals}

In addition to the perspectives discussed above, we can mention three more options for the interpretation of $\psi$ in PWT:

\begin{itemize}
\item [] {\bf Wave Function Realism:}
Here $\psi$ is considered a field in a multi-dimensional configuration space. As \cite{Bell:2004aa}, p.\ 128 claims, the wave function is an objectively real field in configuration space guiding the motion of individual systems. On this view, then, $\mathbb{R}^{3N}$ is the true arena where physics takes place. A version of Bohmian theory defined according to such a perspective is given in \cite{Albert1992}. This view was later refined by \cite{Albert:1996aa} and \cite{Loewer:1996aa}, becoming known as {\em wave function realism}. For a critical assessment of wave function realism, see \cite{Allori:2018}.

\item [] {\bf Functionalism:} 
Following recent research in quantum gravity, where spacetime is defined in terms of its function in the context of the theory, \cite{Allori:2021} extends such an approach to analyze the wave function. According to Allori, $\psi$ is a non-representational entity: the wave function does not represent a physical item in space, nor does it represent an agent's epistemic knowledge. Rather, this concept is directly defined via the roles --- viz.\ functions---that it plays in the context of a certain theory.

\item [] {\bf Multi-Field:} The multi-field account reinstates physical space as the fundamental arena for physics. This perspective interprets $\psi$ as a real multi-field in three-dimensional spacetime, which assigns properties to regions of spacetime, rather than points. As \cite{Chen:2019} suggests, ``we can think of it as a ``multi-field'' that assigns properties to every region of $\mathbb{R}^3$ that is composed of $N$ points''. The central idea of the multi-field approach in pilot-wave theory is to associate a definite field value with $N$-tuples of spatial points corresponding to the exact location of the Bohmian corpuscles. In turn, the particles' coordinates of an actual system ``select'' a unique value for the multi-field in three-dimensional space.\footnote{Objections to the multi-field approach are contained in \cite{Belot:2012aa}.}
\end{itemize}

\section{Pilot-Wave Theory of Quantum Measurements} 
\label{measure}

After having briefly discussed the major formulations of PWT and how Born's rule and $\psi$ are interpreted according to various perspectives, let us now turn to the thorny issue of quantum measurements. Furthermore, in this section, we shall argue that, contrary to many critics of PTW, its structure is simpler with respect to standard QM. 

\subsection{Measurements in PWT}

We have anticipated in Section \ref{Bohm} that \cite{Bohm:1952ab} provided the first theory of quantum measurements, with which he causally explained the actualization of individual experiments' outcomes. Building on this work and subsequent essays by Bell (cf.\ \cite{Bell:2004aa}), D\"urr, Goldstein and Zagnhì developed a theory of measurement for BM (\cite{Durr:2004c}, \cite{Durr:2009fk}, Chapter 9). In what follows, we describe the latter theory, for it is simpler with respect to Bohm's but keeps its essential explanation intact.

To illustrate how the pilot-wave theory of measurement works, let us consider an idealized case of a wave function $\psi$ in a superposition of two states $\psi_1, \psi_2$ corresponding to the possible eigenstates of a two-valued operator $O$, with eigenvalues ``left'' ($L$) and ``right'' ($R$):
\begin{align}
\psi=c_1\psi_{1}+c_2\psi_{2}, 
\end{align}
\noindent where $c_1, c_2\in\mathbb{C}$ and $|c_1|^2+|c_2|^2=1$. Before the measurement's performance, we assume that the macroscopic device used to measure $O$ is in its ready state $\Phi_0$ pointing to a neutral direction, whereas the other admissible pointers' positions will be $\Phi_1$ for $L$ (pointing to the left) and $\Phi_2$ for $R$ (pointing to the right). 

From standard QM, we know that the system and apparatus are initially independent and described by a product wave function. As soon as the measurement begins, the Schr\"odinger dynamics introduces correlations between the system's and the apparatus's wave functions, creating a superposition of disjoint wave packets corresponding to the possible measurement outcomes.  Given the deterministic evolution provided by \eqref{SE}, we obtain a {\em macroscopic} superposition:

\begin{align}
\label{macrosuperpos}
\sum_{i=1,2}c_i\psi_i\Phi_0\longrightarrow\sum_{i=1,2}c_i\psi_i\Phi_i.
\end{align}

Now, in BM a particular measurement situation is formally represented by a couple $(X,Y)\in\mathbb{R}^{3M}\times\mathbb{R}^{3(N-M)}$ where the former variable refers to the actual initial $M$-particle configuration of the subsystem under consideration, and the latter to the actual configuration of particle composing the environment, i.e.\ its complement formed by all the particles not in the subsystem, including the particle configuration of the experimental device which will actually register the measurement result.

We now assume that before the performance of the measurement at time $t=0$, the macroscopic pointer points in the neutral direction, i.e.\ the apparatus is in the ``ready'' state and the system has not yet interacted with it as shown in Figure 2.
\vspace{2mm}

\begin{center}
\includegraphics[scale=0.6]{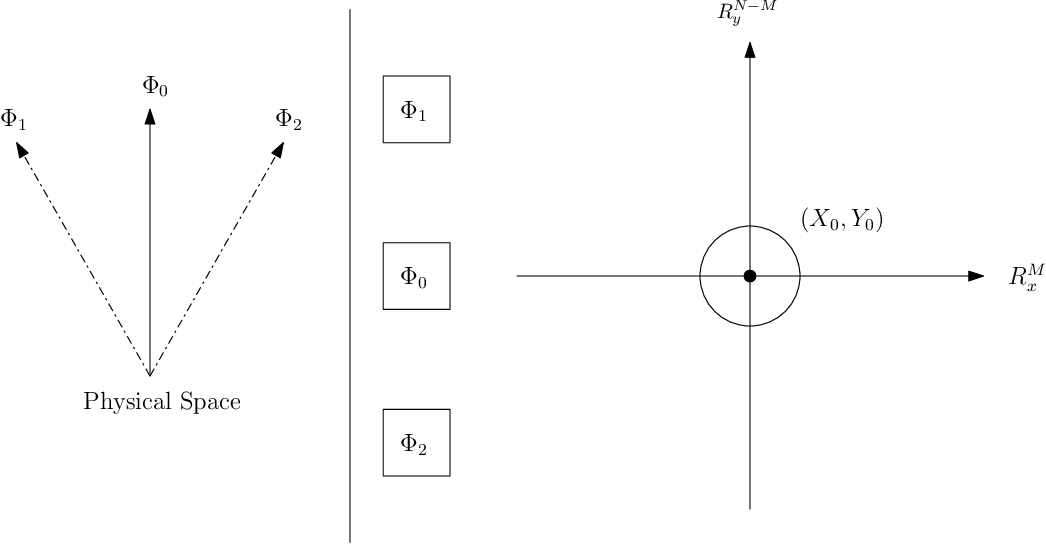}
\begin{quote}
\footnotesize{Figure 2: Schematic representation of the pointer pointing in the neutral direction in physical space before the measurement, solid line. Dashed lines represent physically possible but not yet actualized measurement outcomes (left). The relative support of the pointer's wave function and particle configuration describing the experimental situation before the measurement's performance (right).}
\end{quote}
\end{center}

\noindent As the experiment proceeds, the interaction between the system and the macroscopic device is governed exclusively by the laws \eqref{SE} and \eqref{guide}. Since the particles' positions of the measured system and the experimental apparatus are always well-defined and never in superposition, the measurement outcome is a function of (i) the initial conditions of the particles composing the entire experimental set-up, and (ii) the dynamical laws governing their behavior. 
The initial configuration of particles, in fact, evolves deterministically during the measurement process and enters only one wave packet, remaining there thenceforth. Referring to this, as Passon notes, ``the wavefunction of the measurement apparatus will in general be in a superposition state. The configuration, however, indicates the result of the measurement, which is actually realized. That part of the wavefunction which ``guides'' the particle(s) can be reasonably termed the \emph{effective wavefunction}. All the remaining parts can be ignored, since they are irrelevant to the particle dynamics. As a result of decoherence effects [...], the probability that they will produce interference effects with the effective wavefunction is vanishingly small'' (\cite{Passon:2018}, p.\ 191).\footnote{For a technical discussion on empty branches of the wave function cf.\ \cite{Durr:2009fk}, pp.\ 179-180, and \cite{Bohm:1952aa}, p.\ 182. Additionally, BM provides an algorithm to define the \emph{effective wave function} of the subsystem's degrees of freedom, which corresponds to the usual quantum mechanical wave function. The reader may refer to \cite{Durr:2013aa}, Chapter 2, and \cite{Durr:2009fk}, Chapter 9 for details.} 

The chosen branch of the wave function then guides the particle configuration towards the actualization of a definite state of the macroscopic device, pointing to a given specific direction, actualizing the measurement result. The final configuration $(X_t,Y_t)$, corresponds to one of the possible eigenstates of the measured operator $O$, coupled with a definite state of the macroscopic experimental device, which will point to a definite direction. Thereby, the experimental outcome can be unambiguously read off from the macroscopic measurement device. Suppose now to find the outcome $R$; then the macroscopic pointer will point in the $\Phi_2$ direction, as illustrated in Figure 3.
\vspace{2mm}

\begin{center}
\includegraphics[scale=0.6]{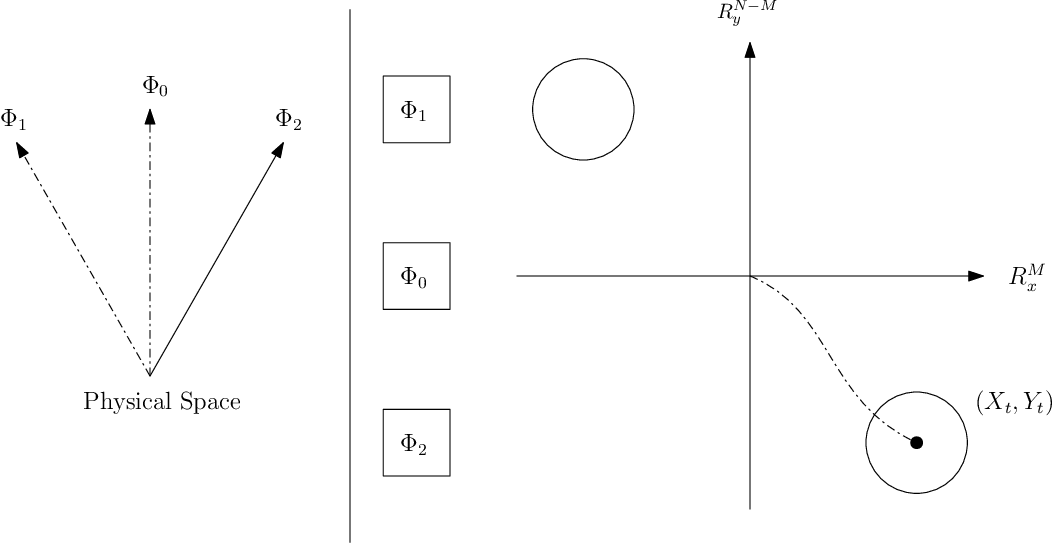}
\begin{quote}
\footnotesize{Figure 3: Schematic representation of the pointer pointing in the right direction in physical space; the outcome $R$ has been obtained, solid line (left). The relative support of the pointer's wave function and particle configuration describing the experimental result is shown on the right. For practical purposes, the empty branch of the wave function can be neglected.}
\end{quote}
\end{center}

\subsection{Remarks on Measurements in PWT}

From our schematic presentation of the Bohmian theory of measurement, a few remarks must be made:

\begin{enumerate}
\item In PWT, there are no superpositions of particles in physical space; therefore, the state described in \eqref{macrosuperpos} is avoided by construction. Consequently, macroscopic superpositions do not occur, solving Schr\"odinger's cat-like scenarios straightforwardly, i.e., the quantum measurement problem\footnote{For detailed accounts of how BM solves the measurement problem, cf.\ \cite{Bricmont:2016aa}, Chapter 5, \cite{Durr:2009fk}, Chapter 9, \cite{Maudlin:1995ab}.};
  
\item A measurement is characterized simply as an interaction between the particles of the observed system and those forming the experimental apparatus. There is no need to introduce the notion of ``measurement'' as an additional primitive concept in the postulates of the theory.\footnote{In the context of Bohm's theory, the explanation of measurement results is structurally the same, but outcomes are derived from the laws \eqref{SE}-\eqref{QHJ}, where the quantum potential plays a crucial role for (i) the formation of wave packets and (ii) the acceleration of particles, acting as a force.} As \cite{Goldstein:2012} underlines, ``a statement like ``the experiment $E$ has the outcome $z$'' should mean that the PO of the apparatus indicates the value $z$. For example, if the apparatus displays the outcome by a pointer pointing to a particular position on a scale, what it means for the outcome to be $z$ is that the matter of the pointer is, according to the PO, in the configuration corresponding to $z$. Thus, the outcome $Z$ is a function of the PO, $Z=f(PO)$'';

\item Similarly, since measurement results are functions of the theory's ontology and its dynamical laws, the physical processes responsible for the macro-objectification of outcomes are independent of external observers, which do not play any active role in the Bohmian theory, contrary to the case of QM;
 
\item Wave functions are not subjected to stochastic jumps: the collapse rule loses its fundamental role in the pilot-wave interpretation, being only an \emph{effect} of the particles' dynamics, thereby dissolving the inconsistency present between the projection rule and the unitary Schr\"odinger equation. Given that PWT dispenses with the collapse rule, which is a theorem in BM, its structure is simpler with respect to that of standard QM, contrary to what is usually affirmed in the literature;
   
\item Neither Bohm nor Dürr, Goldstein and Zanghì aimed at restoring a deterministic picture of physical reality with their interpretations: determinism is indeed inherited from Schr\"odinger's equation, which is the basis of such models;
   
\item Operators are associated with experiments; in PWT, observables do not refer to inherent properties instantiated by a physical system. Indeed, every proposition about quantum measurements must be translated into a statement concerning particles' positions and their dynamics, cf.\ (\cite{Bohm:1952ab}, Section 4, \cite{Durr:2004c});

\item \cite{Oldofredi:2020} shows that the logical structure underlying BM is {\em classical} (cf.\ also \cite{Bacciagaluppi:2009b}). Thus, it does not introduce any modification to standard propositional calculus. In brief, this paper argues that BM implements a classical logical structure at the level of its corpuscular ontology. This fact alone is sufficient to explain why the classical logical connectives maintain their meanings, without the need to define either quantum connectives or alternative interpretations of the usual logical operators. Moreover, it is shown that the distributive law is not violated in the context of BM, providing a powerful counterexample to the belief that we need a new logic to describe the quantum domain. 
\end{enumerate}

\subsubsection{Is PWT Simpler than QM?}

We can positively answer the question asked in the Section title, because we have already seen that the collapse postulate of QM becomes a {\em theorem} of PWT (see remark 4 above), so that the number of postulates decreases in the latter framework---making it formally simpler than standard QM. However, we can say more: one can also argue that the so-called ``additional'' structures of PWT are embedded in the very quantum mechanical formalism and emerge from its fundamental equations. 

Given the above remarks, I propose the following observation: it is misleading to think that PWT implements additional structure---or, as Heisenberg famously claimed, ``ideological superstructure'' (cf.\ \cite{Myrvold:2003})---to the standard quantum formalism.\footnote{Other authors made similar criticisms, e.g., \cite{Wallace:2005}.} If this is correct, then it would be wrong to claim that PWT inflates the structure of QM. 

In support of this observation, I consider a historically relevant case study of Dirac's analogy between classical and quantum mechanics in which he almost derived Bohm's PWT in 1930. Dirac then did not take his analogy to its logical conclusion, for it would allegedly imply a violation of the Heisenberg position-momentum uncertainty relation. In the next Section, we illustrate why Dirac's reasons to discard PWT are unwarranted.
\vspace{2mm}

Let us now focus on paragraph $\S$31 of Dirac's seminal book \emph{The Principles of Quantum Mechanics} (\cite{Dirac:1947aa}). While studying the classical limit of quantum theory, Dirac obtained the equations that constitute the mathematical basis of Bohm's causal approach, as pointed out by \cite{Hiley:2018, Hiley:2019}. Here, it is carefully shown how the classical dynamical equations of Hamiltonian mechanics emerge as a limiting case of QM. Specifically, after having discussed the quantum mechanical treatment of a free particle ($\S$30), Dirac considered the more general case concerning the motion of quantum systems represented by wave packets with positions and momenta having ``numerical values whose accuracy is limited by Heisenberg's principle of uncertainty'' (\cite{Dirac:1947aa}, p.\ 121). It is worth stressing that he analyzed dynamical systems with ``classical analogue'', i.e., those physical systems for which a classical description remains valid as an approximation of a more fundamental quantum mechanical representation.

His analysis began by representing quantum systems in terms of wave functions written in polar form $\psi(q, t)=R(q, t)e^{iS(q, t)/\hbar}$, as Bohm did, where $R, S$ retain the interpretation given in Section \ref{Bohm}. The Schr\"odinger's equation in this case takes the form\footnote{In what follows, we will omit the arguments of $R$ and $S$ to keep the notation as simple as possible. For historical rigor, I employ the same formalism used by Dirac.} 

\begin{align*}
i\hbar\frac{\partial}{\partial t}Re^{iS/\hbar}\rangle=H(q_r, p_r)Re^{iS/\hbar}\rangle,
\end{align*}

\noindent which is equivalent to 

\begin{align}
\label{SE_Dirac}
\left\{i\hbar\frac{\partial R}{\partial t}-R\frac{\partial S}{\partial t}\right\}\rangle=e^{iS/\hbar}H(q_r, p_r)e^{iS/\hbar}Re^{iS/\hbar}\rangle.
\end{align}

\noindent Dirac points out that $e^{iS/\hbar}$ is a unitary operator, that can be used to ``give us a unitary transformation''. Due to the application of such a unitary operator, the $q$ variables remain unchanged, whereas each $p_r$ assumes the following form: 
\begin{align}
\label{H}
e^{iS/\hbar}{p_r}e^{iS/\hbar}= p_r+\partial{S}/\partial{q_r}.
\end{align}

\noindent Now, Dirac underlines that the Hamiltonian appearing in \eqref{SE_Dirac} becomes $e^{iS/\hbar}H(q_r, p_r)e^{iS/\hbar}=H(q_r, p_r+\partial{S}/\partial{q_r})$, so that equation \eqref{SE_Dirac} can be re-expressed as 

\begin{align}
\label{SE_Dirac_2}
\left\{i\hbar\frac{\partial R}{\partial t}-R\frac{\partial S}{\partial t}\right\}\rangle=H\left(q_r, p_r+\frac{\partial S}{\partial q_r}\right)R\rangle.	
\end{align}

Since Dirac aimed to verify that in the classical limit quantum systems obey to classical dynamical laws, he neglected the terms involving $\hbar$ in \eqref{SE_Dirac_2} letting $\hbar\longrightarrow 0$, obtaining that the ``surviving terms'' provide a differential equation for the phase $S$ which is the Hamilton-Jacobi equation (equation (38) in \cite{Dirac:1947aa}, p. 122): 

\begin{align}
\label{S}
-\frac{\partial S}{\partial t}=H_c\left(q_r, \frac{\partial S}{\partial t}\right), 
\end{align}

\noindent where $H_c$ represents the classical Hamiltonian. The British physicist then proceeds to find an equation for the amplitude $R$ of the wave function, which takes the following form (p.\ 123):

\begin{align}
\label{R}
\frac{\partial R^2}{\partial t} = - \sum_s \frac{\partial}{\partial q_s} \bigg[ \frac{\partial H_c (q_r, p_r)}{\partial p_s} \bigg]_{p_r=\partial S/\partial q_r} \bigg\}.
\end{align}

\noindent Here $R^2$ is interpreted as a fluid density, whereas the fluid's velocity is given by
 
\begin{align}
\label{43}
\frac{d q_s}{d t} = \bigg[ \frac{\partial H_c (q_r, p_r)}{\partial p_s} \bigg]_{p_r=\partial S/\partial q_r}.
\end{align}

\noindent Equation \eqref{R} is then the conservation equation for the fluid, whose motion is determined by the phase $S$ satisfying \eqref{S}. Now, Dirac emphasizes that 

\begin{quote}
For a given $S$, let us take a solution of \eqref{R} [equation number and symbols adapted] for which at some definite time the density $R^2$ vanishes everywhere outside a certain small region. We may suppose this region to move with the fluid, its velocity at each point being given by \eqref{43}, and then the equation of conservation \eqref{R} will require the density always to vanish outside the region. [...] We thus get a wave function representing a state of motion for which the coordinates and momenta have approximate numerical values throughout all time. Such a state of motion in quantum theory corresponds to the states with which classical theory deals. The motion of our wavepacket is determined by equations \eqref{S} and \eqref{R} (\cite{Dirac:1947aa}, p.\ 124).
\end{quote}

As correctly underlined in \cite{Hiley:2018, Hiley:2019}, Dirac did not pursue his analogy between the classical and the quantum case any further---i.e., he did not retain all the powers of $\hbar$ as done in \cite{Bohm:1952aa}---since the quantum formalism forbids the simultaneous assignment of well-defined values for positions and momenta to quantum particles in virtue of the Heisenberg principle:

\begin{quote}
By a more accurate solution of the wave equation, one can show that the accuracy with which the coordinates and momenta simultaneously have numerical values cannot remain permanently as favourable as the limit allowed by Heisenberg's principle of uncertainty [$\dots$] (\cite{Dirac:1947aa}, p.\ 125).
\end{quote}

Bohm's approach, on the contrary, pushes Dirac's analysis to its logical conclusion since he noticed that a ``classical'' description of physical systems, i.e., a description where precise values for position and velocities of quantum objects are always definite, is neither in contradiction with the formal structure of QM---in particular with its non-commuting algebraic structure as we aregoing to see later---nor with the position-momentum uncertainty relation (even when higher orders of $\hbar$ are retained).

Referring to this, \cite{Hiley:2018} derived Bohm's theory from Dirac's approach, explaining how the causal approach can be straightforwardly obtained from QM. 
Hiley and Dennis begin their discussion about Dirac's approach to quantum theory from the relation \eqref{H}, encountered a few lines above. First, the authors plug the Hamiltonian $H(\hat{q}, \hat{p})=e^{iS(q,t)/\hbar}H(\hat{q}, \hat{p})e^{iS(q,t)/\hbar}$ into the usual Schr\"odinger's equation, which is the starting point used by both Dirac and Bohm. After having decomposed the wave function in polar form $\psi(q, t)=R(q, t)e^{iS(q, t)/\hbar}$, the authors re-express \eqref{SE} as follows:

\begin{align}
\label{SE_2}
i\hbar R(q, t)^{-1}\frac{\partial{R(q, t)}}{\partial{t}} -\hbar\frac{\partial{S(q, t)}}{\partial {t}}=R^{-1}(q, t)H R(q, t).
\end{align}

\noindent In this way, it is possible to divide it into its real and imaginary parts.\footnote{To write them down Hiley and Dennis split also $\hat{p}^2$ in its imaginary and real parts, which can be respectively written as follows: $\Re(\hat{p})= \hat{p}^2 +\left(\frac{\partial S}{\partial q} \right)^2$ and $\Im (\hat{p})= -i\hbar\frac{\partial^2{S}}{\partial{q^2}}+2\left(\frac{\partial S}{\partial q} \right)\hat{p}$.} For simplicity they choose a Hamiltonian of the usual form $H=\hat{p}^2/2m + V$, showing that the real part

\begin{align}
\label{Dirac_Bohm}
\frac{\partial S(q, t)}{\partial t} + \frac{1}{2m}\left(\frac{\partial S(q, t)}{\partial q}\right)^2 - \frac{\hbar^2}{2mR(q, t)}\left(\frac{\partial^2R(q, t)}{\partial q^2}\right)+V(q)=0
\end{align}
\noindent is precisely equation \eqref{QHJ} obtained by Bohm (cf. \cite{Bohm:1952aa}, p.\ 170, Eq.\ 6). On the other hand, the imaginary part of the Schr\"odinger's equation becomes

\begin{align*}
i\hbar\frac{\partial R(q, t)}{\partial t} = \left[-i\hbar\left(\frac{\partial^2 S(q, t)}{\partial q^2}\right) +2\left(\frac{\partial S(q, t)}{\partial q}\right)\hat{p}_S\right]R(q, t).
\end{align*}

\noindent Imposing that $\rho(q, t)=R^2(q,t)$, the latter equation clearly expresses the conservation of probability if it is rewritten as follows:

\begin{align}
\label{Dirac_Bohm_2}
\frac{\partial \rho(q, t)}{\partial t} +\frac{1}{m}\frac{\partial}{\partial q}\left(\rho(q, t)\frac{\partial S(q, t)}{\partial q}\right)=0,
\end{align}

\noindent it is important to note that \eqref{Dirac_Bohm_2} is equivalent to equation \eqref{density} (eq.\ 5 in  \cite{Bohm:1952aa}, p.\ 170). Therefore, starting from the Schr\"odinger's equation, Hiley and Dennis showed how to arrive at the defining equations of Bohm's theory \eqref{density}, \eqref{QHJ} from Dirac's approach.\footnote{More recently, \cite{Sakurai1994} pp.\ 102-103 derived \eqref{Bv} from \eqref{SE} writing the wave function in polar form. As in Dirac's case, Sakurai also did not interpret \eqref{Bv} as a particle velocity, for it would contradict the Heisenberg position-momentum uncertainty relation.} In this regard, they concluded that 

\begin{quote}
the Dirac approach and the Bohm approach lead to the same equations. For this reason, we will call the representation the Dirac-Bohm picture (\cite{Hiley:2018}, p.\ 3).
\end{quote}

\noindent Furthermore, it is very important for our discussion to stress that 
\begin{quote}
Dirac remarked that equation \eqref{Dirac_Bohm_2} [numeration adapted] was similar to the continuity equation used in fluid dynamics but did not mention the earlier work of Madelung [$\dots$]. he further noted that equation \eqref{Dirac_Bohm} [numeration adapted] had the form of the classical Hamilton-Jacobi equation if one neglected terms of order $\hbar^2$ and above. Dirac did not pursue this line of reasoning because he thought it would come into conflict with the uncertainty principle (\cite{Hiley:2018}, p. 4).
\end{quote}

\subsubsection{Discussion}

The previous discussion showed that Dirac could have derived the equations defining Bohm's theory from Schr\"odinger's equation. This fact clearly indicates that the ``additional structure'' usually ascribed to his interpretation is instead rooted in the formal structure of standard QM. As we have seen, however, Dirac did not pursue a complete analogy between classical and quantum mechanics because of an apparent conflict with the Heisenberg uncertainty principle. 
Nonetheless, \cite{Bohm:1952ab} provides compelling reasons as to why Dirac's belief is ill-founded. 

To show that, it is sufficient to take into account the theory of measurement introduced in this Section. In PWT, particles' positions are not known, so one can say that the particles are somewhere in those regions where $|\psi|$ is appreciable. Moreover, the momentum of a particle in Bohm's theory equals $\nabla S(x)$; however, since $x$ is unknown, the momentum is generally not specified. Referring to this, Bohm claims that 
\begin{quote}
as long as we are restricted to making observations of this kind, the precise values of the particle position and momentum must, in general, be regarded as ``hidden,'' since we cannot at present measure them. They are, however, connected with real and already observable properties of matter because (along with the $\psi$-field) they determine in principle the actual result of each individual measurement (\cite{Bohm:1952ab}, p.\ 183). 
\end{quote}

\noindent This is what \cite{Bell:2004aa} and \cite{Goldstein:2025} mean when they say that ``hidden variables'' is an unhappy terminology, for measurement results are seen in configurations of hidden variables which constitute microscopic and macroscopic objects, not in the wave function.\footnote{\cite{Bell:2004aa}, p.\ 201, writes that ``Absurdly, such theories are known as ``hidden variable'' theories. Absurdly, for there it is not in the wavefunction that one finds an image of the visible world, and the results of experiments, but in the complementary ``hidden''(!) variables. Of course, the extra variables are not confined to the visible ``macroscopic'' scale. For no sharp definition of such a scale could be made. The ``microscopic'' aspect of the complementary variables is indeed hidden from us. But to admit things not visible to the gross creatures that we are is, in my opinion, to show a decent humility, and not just a lamentable addiction to metaphysics. In any case, the most hidden of all variables, in the pilot wave picture, is the wavefunction, which manifests itself to us only by its influence on the complementary variables''.} Even though, strictly speaking, in PWT one asserts that particles have precise momenta and positions and they explain measurement outcomes, these variables are unknown and uncontrollable, so that for practical purposes one needs to ``restrict oneself to a statistical description of the connection between the values of these variables and the
directly observable results of measurements. Thus, we are unable at present to obtain direct experimental evidence for the existence of precisely definable particle positions and momenta'' (\cite{Bohm:1952ab}, p.\ 183). In virtue of this feature of PWT, Heisenberg's position-momentum uncertainty relation is not violated. 

However, another, deeper, principled reason grounded in the Bohmian theory of measurement can be given. In this perspective, during a measurement of the observable $X$---denoting here the position operator---the interaction between the measured particle and the experimental apparatus transforms $\psi$ from its initial state (whatever it may be) into an eigenfunction of said operator, $\psi_x(q)$. The value of the measurement outcome is random, for its initial conditions are neither known nor controllable. Now, if we measure the momentum observable $P$, which does not commute with $X$, the result of such a measurement will be random as well. The initial wave function, which is in an eigenstate of the position operator, now transforms into another state, which is an eigenstate of $P$, with an associated eigenvalue $p$ which represents ``the packet actually entered by the apparatus coordinate. Appealing to the algebraic structure of QM, Bohm concludes that even in PWT, one cannot simultaneously measure incompatible observables on the same system, for each measurement

\begin{quote}
disturbs the system in a way that is incompatible with carrying out the process necessary for the measurement of the other. Thus, a measurement of $P$ requires that wave field, $\psi$, shall become an eigenfunction of $P$, while a measurement of $X$ [notation adapted] requires that it shall become an eigenfunction of $X$. If $P$ and $X$ do not commute, then by definition, no $\psi$-function can be simultaneously an eigenfunction of both. In this way, we understand in our interpretation why measurements, of complementary quantities, must (as in the usual interpretation) necessarily be limited in their precision by the uncertainty principle ({\em ibid}.).
\end{quote}

In sum, Heisenberg's uncertainty principle is fully compatible with PWT, so that Dirac's (and Sakurai's) conclusions are unwarranted. Therefore, the quantum formalism itself contains the tools to define particles' velocity in a way that does not generate any inconsistency with, or violation of, the position-momentum uncertainty relation. 

If our argument is correct, then PWT does not add structures to the formalism of QM. Thus, the critics of this approach should not appeal to the metaphysical virtue of simplicity to argue for the superiority of standard QM over PWT. On the contrary, supporters of the latter framework should employ simplicity-based arguments to defend it. 

\section{On the Nature of Spin}
\label{Spin}

\subsection{Contextuality and Spin in Bohmian Mechanics}
\label{nospin}

The previous Section illustrated that in PWT measurements of quantum observables are reduced to position measurements. It is not surprising, then, that such a framework is {\em contextual}. On the one hand, outcomes are ontologically dependent on the initial conditions of the experimental set-up and the dynamics of the theory. On the other hand, quantum observables are not metaphysically genuine properties of the particles; consequently, their measurements do not reveal pre-existing values of such magnitudes (cf.\ \cite{Bohm:1952ab}, Section 4, \cite{Durr:2004c}). 

Referring to this, many authors argued in particular that in BM spin does not exist (cf.\ \cite{Bell:1982}, \cite{Bricmont:2018}, \cite{Durr:2004c}, \cite{Durr:2009fk}). Let us review their common view with an example, taking into account a spin measurement on a single particle along the $z$-axis.\footnote{This example is taken from \cite{Bricmont:2018}.} 

Before the measurement of the $z-$spin, suppose that the following state describes the system:
\begin{align}
\psi(z)\Big(|\uparrow\rangle+|\downarrow\rangle\Big).
\end{align}

For simplicity, we assume that the wave function written above is symmetrical, i.e.\ $\psi(z)=\psi(-z)$, so that the nodal line shown in Figure 4 corresponds to the case in which $z=0$; such a nodal line cannot be crossed by Bohmian particles in virtue of \eqref{guide}. In the figure below, $H$ represents the inhomogeneous magnetic field of the idealized Stern-Gerlach apparatus we are considering, whereas the arrow indicates its direction; the circles represent the supports of the wave function. 
\vspace{4mm}

\begin{center}
\includegraphics[scale=0.6]{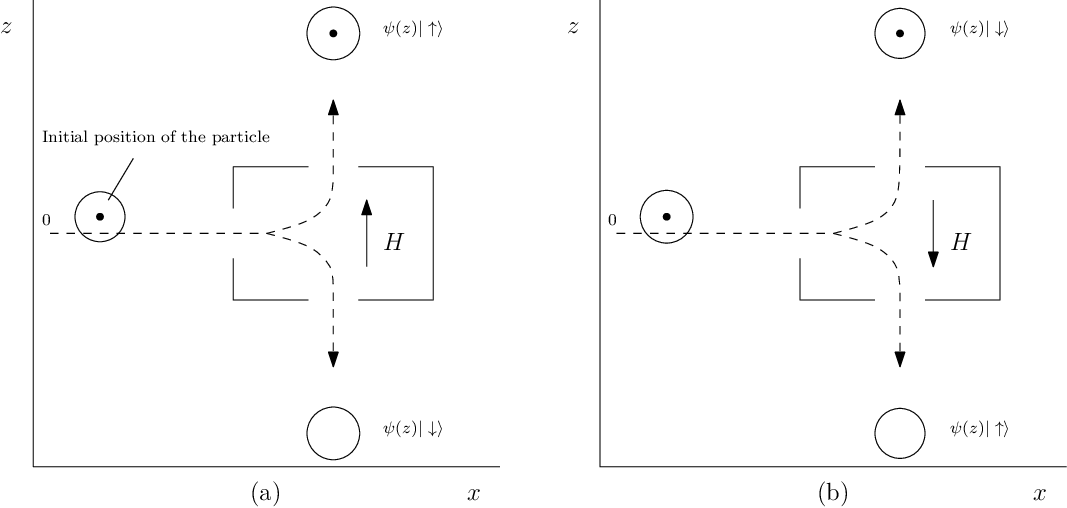}
\begin{quote}
\footnotesize{Figure 4: Schematic representation of the contextual nature of the property of spin in BM.}
\end{quote}
\end{center}

Now, performing a $z$-spin measurement, one can obtain two results: $z$-spin-up ($|\uparrow\rangle$), which in Figure 4a follows the direction of $H$, and $z$-spin-down ($|\downarrow\rangle$), which goes in the opposite direction. In our example, at the beginning of the measurement, the Bohmian particle is located above the nodal line, so it will necessarily go up, obtaining the state $\psi(z)|\uparrow\rangle$. We therefore say that the particle has ``$z-$spin up''. Reversing the direction of $H$, but leaving unaltered the initial position of the particle (Figure 4b), we conclude that the corpuscle will go up as well. However, this time the particle travels in the opposite direction with respect to the $H$ field; thus, the particle has ``$z-$spin \emph{down}''. 
Since in BM spin measurements are entirely explained in terms of the initial particle configuration of the experimental context--- together with the dynamics of the theory---it follows that an observation of ``spin'' does not measure any pre-existing property of the particle under consideration. Consequently, adopting an eliminative form of reductionism---typical of the PO programme---it is concluded that spin does not exist.

The reader will notice here an application of an indispensability-type of argument, which is employed by the PO perspective, as claimed in \cite{Allori:2018}. According to Allori, taking into account a PO theory, one should be committed only to those entities indispensable for the explanation of a certain physical phenomenon---the primitive variables. In this case, thus, particle positions are ontologically salient, whereas spin is not. Given that the latter notion is dispensable, one should not be committed to the existence of this property for the Bohmian corpuscles.

\subsection{A Consequence for Electrons' Trajectories in the Hydrogen Atom}
\label{teufel}

A striking consequence of the non-existence of spin in BM appears if we look at the trajectory of electrons guided by a ground state wave function of a hydrogen atom.  

\cite{Durr:2009fk} affirms that the ground state of the hydrogen atom can always be chosen real and everywhere positive. This fact has remarkable implications for BM: since the guiding field is real, it entails that the imaginary part of the vector velocity field appearing in the guiding equation \eqref{guide} is zero. This implies that the electron's velocity orbiting the hydrogen nucleus is zero, so the electron is at rest in the ground state of the hydrogen atom. 

Referring to this peculiar feature of BM, Dürr and Teufel claim that this is a radical innovation---contrary to the naive conception of Bohr according to which electrons do circulate the proton of the hydrogen atom---and one should accept even counter-intuitive aspects of the theory insofar as they are mathematically and empirically consistent. 
Conforming to the Bohmian treatment of the electron's trajectories in the ground energy state of the hydrogen atom, we can only say that the various possible positions of the electron around the nucleus are $|\psi_0|^2$-distributed in virtue of the quantum equilibrium hypothesis. This fact in turn implies that if we consider an ensemble of hydrogen atoms in the ground energy state ``the empirical distribution of $Q$ is typically close to $|\psi_0|^2$. That can be checked by experiment. The spread in position is then simply the variance of the $|\psi_0|^2$-distribution'' (\cite{Durr:2009fk}, p.\ 154), as we can see from Figure 5 below. 
\vspace{2mm}

\begin{center}
\includegraphics[scale=0.4]{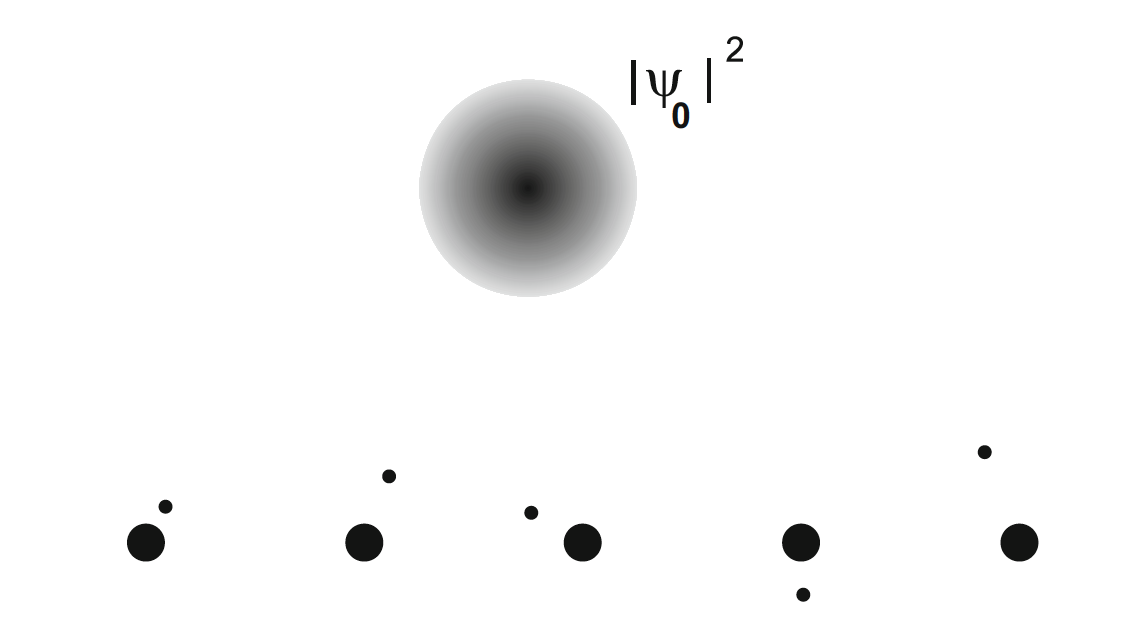}
\begin{quote}
\footnotesize{Figure 5: Pictorial representation of the ground-state wave-function distribution for an ensemble of hydrogen atoms in the ground state. Here, electrons' positions are $|\psi_0|^2-$distributed. This picture is taken from \cite{Durr:2009fk}, p.\ 154.}
\end{quote}
\end{center}

\noindent In this regard, Norsen maintains that:
\begin{quote}
The electron in a Hydrogen atom in its ground state is not, according to the pilot wave theory, orbiting the proton, but is instead just sitting there, at some fixed point near the proton (\cite{Norsen:2017}, p.\ 183).
\end{quote}

\noindent A few lines below, he makes another substantial remark:
\begin{quote}
[i]f that bothers you or seems physically impossible, you are probably tacitly expecting that if the electron is literally a particle, it should obey Newton's equations of motion, and should therefore accelerate toward the proton due to the electrostatic force. But the pilot-wave theory is not classical mechanics! The motion of the particle, according to this theory, is not determined by classical forces acting on it, but is instead determined by the structure of the wave function which guides it (\emph{ibid.}).
\end{quote}

Notwithstanding the mathematical coherence of BM and its non-classicality underlined by Dürr, Teufel, and Norsen, there are pilot-wave theories in which spin is considered a real degree of freedom, and this fact entails notable physical consequences. In what follows, in fact, we see that considering spin-dependent trajectories, the pilot wave theory tells a different story about the electron's orbit in the hydrogen atom's ground energy state. To this we now turn. We shall discuss the metaphysical implications of these models later in this Section.

\subsection{Alternative Approaches to Spin in Pilot-Wave Theory}
\label{Bohm_spin} 

As soon as Bohm's 1952 papers were published, he worked on extensions of his PWT. In \cite{Bohm:1953aa}, the causal approach was applied to the Dirac equation, while in 1955, he addressed the question of the nature of spin, employing the Pauli equation as the foundation of particle dynamics (\cite{Bohm:1955a}, \cite{Bohm:1955b}). 

In these essays, Bohm and his collaborators began their analysis with a hydrodynamic model similar to that studied in \cite{Bohm:1954aa}, extending it to describe a spinning electron. They firstly considered the Pauli equation written as follows:
\begin{align}
\label{Pauli}
i\hbar\frac{\partial \Psi}{\partial t}=\bigg(-\frac{\hbar^2}{2m}\bigg(\nabla-ie\frac{\mathbf{A}}{c}\bigg)^2 + V + \frac{e\hbar}{2mc}(\mathbf{\sigma \cdot B})\bigg)\Psi,
\end{align}

\noindent where $\Psi$ is a two component spinor, $\mathbf{A}$ is the electromagnetic vector potential, $V$ total potential energy and $\mathbf{B}$  the magnetic field. Secondly, they define both the fluid density $\rho=\Psi^*\Psi$ and the fluid velocity. The latter is written as the ratio between the conserved charge ($\rho=\Psi^*\Psi$) and conserved current ($\mathbf{j}= \frac{\hbar^2}{2mi}(\Psi^*\nabla\Psi-\Psi\nabla\Psi^*) - \frac{e}{c} \mathbf{A}$) of the Pauli equation:

\begin{align}
\mathbf{v} = \frac{\hbar^2}{2mi}\frac{(\Psi^*\nabla\Psi-\Psi\nabla\Psi^*)}{\Psi^*\Psi} - \frac{e}{c} \mathbf{A}. 
\end{align}

\noindent Having defined the basic elements of the model, the authors assume a new physical property {\em of the fluid}, affirming that it possesses an intrinsic angular momentum, i.e.\ spin, which is obviously connected with the spinorial elements of $\Psi$. 

In this theory, spin assumes a very clear meaning, being the continuous rotational motion of the very small rigid bodies composing the fluid. For instance, to say that a certain particle is in a ``spin up'' state along the $x$-axis means that such a particle has a spin vector---specified in terms of Euler angles describing the orientation of a rigid body with respect to a set of coordinates, and which constitute a new set of hidden parameters---which points along that axis and makes the particle moving in a certain way, whereas the $y$ and $z$ components are not indetermined but {\em zero}---for the particle is not rotating along those other axes.

Finally, the authors underline that the orientation of such spinning bodies depends on $\Psi$. Since the causal interpretation of the Pauli equation generalizes Bohm's previous work on the Schr\"odinger equation---where the wave function $\psi$ was interpreted as a real physical field---they wanted to provide a method to interpret all the four real components of the spinorial wave function ``in a visualizable way by constructing a mathematically equivalent set of fields in euclidean space'', as explained in \cite{Dewdney:1988}. 

Such spinorial wave function is indeed interpreted as a multi-component field that guides the motion of a rigid spinning body, whose centre of mass follows a deterministic trajectory in space generally different from the trajectories implied by the Schr\"odinger equation, for a spin dependent force appears in the quantum potential contained in the generalized quantum Hamilton-Jacobi equation obtained from the Pauli dynamics (for more technical details cf.\ \cite{Bohm:1955a}, p.\ 63, \cite{Dewdney:1988}, pp.\ 537-538, \cite{Holland:1993}, chapters 9 and 10).

Another interesting metaphysical feature of this model---unfortunately often overlooked in philosophical discussions---is that the authors explicitly claim that particles have a small but non-zero size. Hence, they do not approximate quantum particles to mathematical points, as instead done in the original 1952 papers, as well as in other works in BM (\cite{Durr:2013aa}, \cite{Goldstein:2005b, Goldstein:2005a}, \cite{Esfeld:2017}). As Bohm later remarked, in the causal interpretation of the Pauli equation ``the bodies with which we are concerned are not points'' (\cite{Bohm:1957}, p.\ 81).\footnote{A detailed discussion on the notion of particle in pilot-wave theory is given in \cite{Oldofredi:2025a}.}

Further developments of pilot-wave theory saw new ideas concerning the status of spin, as exemplified by the already mentioned works of \cite{Dewdney:1988} and \cite{Holland:1993, Holland:1999}. 

In agreement with Bohm, Schiller, and Tiomno, these authors claim that spin is a real, continuous, and deterministic vector; however, these works provide a particle model, thereby offering a framework closer to the usual formulations of pilot-wave theory as defined in Section \ref{Bohm}.\footnote{Bohm later rejected his 1955 works on spin: in the last part of his career he defined spin as a component of the wave function, i.e.\ not a property of the quantum corpuscles, cf.\ \cite{Bohm:1993aa} for details.}  

Referring to this, they suppose that a ``spinor wavefunction defines a physically real multi-component field whose structure may be understood in terms of the interconnected functions $\rho(=R^2)$, $S$ and ${\bf S}$, or $\rho, \theta, \psi$ and $\chi$'', the variables specifying the Euler angles describing the particles' vectorial orientation (\cite{Dewdney:1988} p.\ 537). The wave field, they say, ``enters into potential, force, and torque terms in the equation of motion and precession of the spinning body'', in analogy with Bohm's 1952 theory. Moreover, $\rho$ plays the usual role of a probability density of a given particle configuration, whereas the spin vector ${\bf S}$ ``has the role of the Hamilton-Jacobi action function for the translational motion of the particle'' (\emph{ibid.}). Finally, Dewdney {\em et al.}, preserved the interpretation of spin as the rotational motion of non-zero-sized particles, for they claim that ``[i]n addition to this wave there is a small rigid spinning body which moves within the wave and is guided by it. The centre of mass of the body pursues a track $x=x(t)$ and we suppose that its internal angular momentum vector is just the spin vector field ${\bf S}$ evaluated along its trajectory'' (\emph{ibid.}). 

Hence, the non-commutativity of spin orientations assumes a clear meaning: if a given corpuscle is rotating along a given axis, it cannot be rotating along different axes for which the value of spin will be null. This fact, in turn, avoids interpretational issues concerned with the metaphysical indeterminacy of the spin property present in QM.\footnote{For a detailed examination of metaphysical indeterminacy in PWT, the reader may refer to \cite{Lewis:2016aa} and \cite{Oldofredi:2024}.}

As we are going to see, such models---and with them, their conception of spin---proved to be relevant for contemporary research in pilot-wave theory. 

\subsection{A New Perspective on the Hydrogen Atom Ground State}
\label{CV}

Building on \cite{Holland:1993, Holland:1999}\footnote{Chapters 9 and 10 of \cite{Holland:1993} extend and systematize the ideas contained in \cite{Dewdney:1988}, a paper that Holland coauthored.}, \cite{Colijn:2002, Colijn:2003} explicitly show that taking into account spin-dependent trajectories for Bohmian corpuscles, one can avoid unwelcome consequences such as those illustrated in Section \ref{teufel}.

Colijn and Vrscay claim that equations \eqref{SE} and \eqref{guide} are relevant for spin-0 particles, since they do not include spin components. For particles with spin, however, these equations ``are inconsistent if the theory is to be ultimately embedded in a relativistic theory'' (\cite{Colijn:2002}, p.\ 334). Moreover, as pointed out in \cite{Holland:1999}, Lorentz symmetry requires that the momentum for a particle with spin $\textbf{s}$ is given by\footnote{Following Bohm's 1952 papers, \cite{Colijn:2002} consider a wave function written in polar form, i.e.\ $\psi(x,t)=R(x,t)e^{iS(x,t)/\hbar}$ where $S$ is the phase of the wave function, the momentum for spin-less particles is given by $\textbf{p}=\nabla S$.}:
\begin{align}
\label{spin}
\textbf{p}=\nabla S + \nabla \textrm{log} \rho \times \textbf{s},
\end{align}

\noindent where $\rho=\psi^*\psi$, the current vector associated with the above equation has the following form:
 \begin{align}
\label{current}
\textbf{J}=\frac{1}{m}\rho\nabla S + \frac{1}{m}\nabla \textrm{log} \rho \times \textbf{s}.
\end{align}

In this framework, $\textbf{J}$ is the Pauli current and explicitly accounts for the spin components of Bohmian particles---note that in \cite{Colijn:2002} spin coordinates are included in the wave function, and particles are treated as points, recalling Bohm's 1952 theory. Now, Colijn and Vrscay underline that if the momentum is defined by equation $\textbf{p}=\nabla S$, then electrons are at rest in the Hydrogen ground state, for $\nabla S=0$, in agreement with the result mentioned in Section \ref{teufel}. 

However, including spin components within the definition of the Bohmian particles' momentum, as in \eqref{spin}, such a counterintuitive feature of BM {\em vanishes}, since now a complete representation of the electron in the Hydrogen atom must include both spatial and spin information. 
Let's illustrate a simple example by considering an electron with spin vector $\textbf{s}=\hbar/2\textbf{k}$ that is in a hydrogenic eigenstate $\psi(x, 0)=\psi_{nlm}(x)$. As Colijn and Vrscay note, the dynamical evolution of such a wave function is given by 
\begin{align}
\psi(x, t)=\psi_{nlm}(x)e^{-iE_nt/\hbar}.
\end{align}

\noindent Since we consider the momentum of the particle as defined by equation \eqref{spin} and in this case $\nabla S=0$, it follows that the momentum of the electron this time is not zero, but is given by the spin term $\nabla \textrm{log} \rho \times \textbf{s}$. This fact, in turn, entails several interesting consequences: 

\begin{quote}
First, the vector $\nabla \textrm{log} \rho$ points in the direction of the steepest increase in $\textrm{log} \rho$, hence in $\rho=|\psi|^2$. Because of the cross product, the momentum vector \textbf{p} is perpendicular to this direction. In other words, the trajectories of the electron lie on level surfaces of $|\psi|^2$. However, \textbf{p} is also perpendicular to the direction of the spin, assumed to lie along the $z$-axis in this discussion. This implies that $z$ is constant for these Bohm trajectories. From the above analysis, the shape of the Bohm trajectories may be found by computing level surfaces of $|\psi|^2$---or simply $\psi$ for real-valued eigenfunctions---and then finding the intersections of these surfaces with planes of constant $z$ (\cite{Colijn:2002}, p.\ 335-336).
\end{quote}

\noindent From their analysis, the authors conclude that an electron in the 1$s$ state of the Hydrogen atom must execute circular motions around the $z$-axis. Hence, Colijn and Vrscay showed that, contrary to the case of spinless particles considered in BM, where particles are stationary and at rest, the spin-dependent trajectories for ground---and also for excited states---of the Hydrogen atom are stable periodic orbits.
Referring to this, \cite{Colijn:2003}, taking into account spin-dependent trajectories,  describe the transition from state 1s to state 2$p_0$ as a consequence of the action of an oscillating electron field.

\begin{center}
\includegraphics[scale=.50]{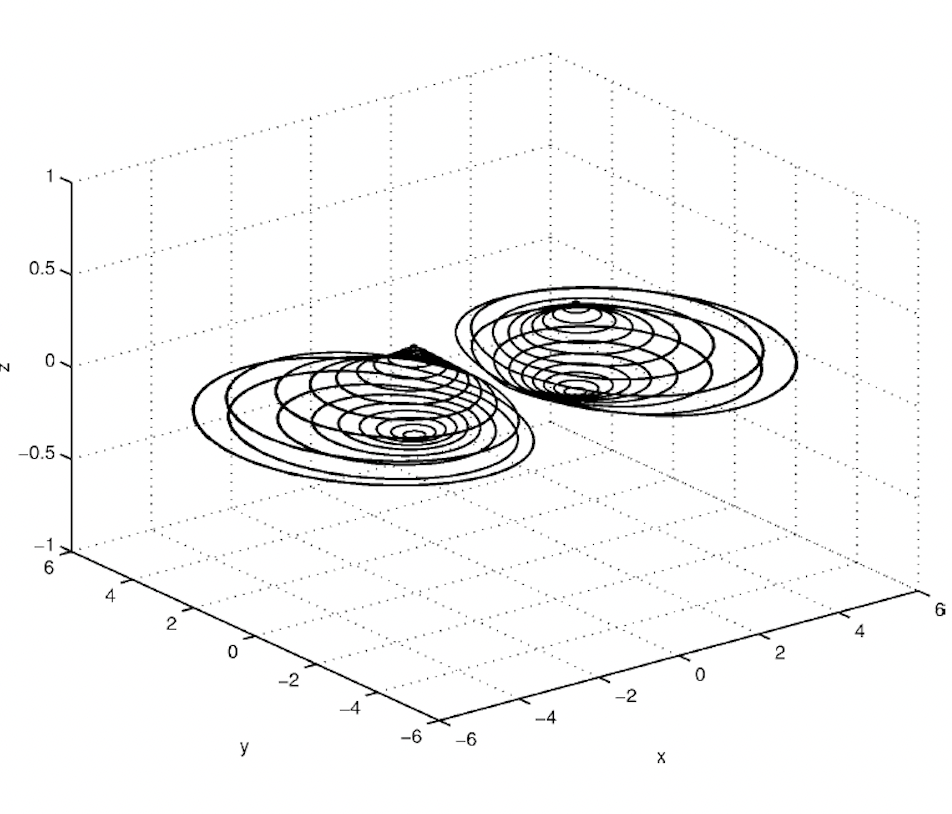}

\small{Figure 6: Spin-dependent trajectories for the 2$p_x$ hydrogen eigenstate. 

Picture taken from \cite{Colijn:2002}.}
\end{center}

The models proposed by Colijn and Vrscay match chemists' way of conceptualizing molecular structures, which describe electrons orbiting around nuclei, and have tools to describe transitions to higher energy states. Indeed, as pointed out by \cite{Fortin:2025}, p.\ 257, ``the image of electrons at rest in atoms and molecules looks strange to the eyes of chemists, accustomed to think of electrons somehow `orbiting' around nuclei''. Thus, they conclude that ``these results are intuitively more acceptable than the original Bohmian results that {\em all} trajectories associated with a given eigenstate are stationary'' (\cite{Colijn:2002}, p.\ 338).

\subsection{Spin-Dependent Trajectories for Time-of-Flight Measurements}

In addition to the frameworks discussed in the previous Section, another interesting application of BM with spin-dependent trajectories is the description of time-of-flight measurements (\cite{Das:2019, Das:2025}).\footnote{Since they discuss spin-$1/2$ particles, the usual laws \eqref{SE}-\eqref{guide} prove to be inadequate, as already pointed out above.} 

It is well-known that in QM there's no time operator; thus, to predict {\em when} a certain particle reaches a detector is a non-trivial issue for the standard theory. Since in the pilot-wave approach the notion of trajectory is embedded, Das uses this tool to compute unambiguous time-arrival distributions, a major theoretical advancement with respect to orthodox QM. Indeed, as Das and D\"urr underline, ``[t]he random first arrival time $\tau$ is then simply the time at which the random Bohmian trajectory $R(t, R(0))$ strikes the surface'' of a detector---it is worth noting for our discussion that the spin compoenent of the Bohmian velocity heavily influence and gives form to particles paths, and thereby to arrival time distribution (in these essays spin is not a property of the quantum corpuscles; rather, the authors employ a spinorial wave function as Colijn and Vrscay did).

Moreover, \cite{Das:2019} devises a potential time-of-flight experiment to test (or falsify) BM, since in this scenario pilot-wave theory predictions differ from those of QM. In the ideal experiment proposed in \cite{Das:2019}, particles are prepared in various spin-polarized states $|\Psi_i\rangle=|\psi\rangle \otimes |\hat{\textbf{n}}\rangle$ (being $\hat{\bf n} \in \mathcal{S}^2$ the direction of spin-polarization) with identical spatial states $|\psi\rangle$.\footnote{This essay generated heated discussions among Bohmians, indeed, recently \cite{GTZ:2024} argued that the trajectories computed by Das and D\"urr would be non-measurable. In a nutshell, Goldstein {\em et al.} claim that ``[t]he probability distributions they [Das and D\"urr] found depend upon the spin of the particle and for some choices of the spin the distribution exhibits intriguing characteristics. They assumed that the arrival times could be measured using appropriate detectors''. Such an assumption, it is claimed, would be false, for no such detector can exist. This conclusion holds because the statistics of the envisioned experiment cannot be given in terms of a positive-operator-valued measure (POVM). Since it is a theorem of BM that ``[t]he statistics of the outcomes of any quantum experiment are governed by a POVM'', then no such experiment can exist (cf.\ \cite{Beck:2025}). \cite{Das:2023} replied by questioning the criterion for measurability employed by Goldstein {\em et al.}, which would not be suitable for ``judging with certainty what can or cannot be observed in actual experiments''. Moreover, they claim that ``the arrival-time problem has long been challenging the ``all's POVM postulate of standard QM. Among the many proposed solutions, some comply with it and some do not''. As the present paper is being written, this dispute remains open.}

\subsection{Discussion}

The argument offered in Section \ref{nospin} rules out the existence of spin by showing that spin measurements reduce to position measurements; therefore, contrary to standard QM, spin is not considered a genuine property of the Bohmian particles. It is certainly true that in any Bohmian theory the spin observable does not refer to any quality instantiated by corpuscles; hence, in this precise sense, it is correct to claim that {\em qua inherent attribute of the particle} it does not exist. 

However, this conclusion should not preclude the possibility of defining spin in a different way with respect to standard QM and including this new characterization of it in PWT, assigning it ontological commitment. In the models introduced in Sections \ref{Bohm_spin}, \ref{CV}, in fact, spin was not defined as an observable representing an intrinsic property of the corpuscles. Rather, it was defined as a degree of freedom influencing the dynamical behavior of the particles, i.e., a further parameter in addition to position needed to fully describe the state of motion of Bohmian corpuscles. 

Colijn and Vrscay's analysis of the ground state of the hydrogen atom clearly shows that the information deriving from spin is not reducible to spatial information alone. It is the spin that crucially explains and describes the particle's orbit around the hydrogen nucleus. Therefore, it becomes essential to avoid the unphysical behavior entailed by standard BM seen in Section \ref{teufel} (cf.\ \cite{Fortin:2025}). Similarly, it is the spin component of the Bohmian trajectories that is essential to compute the arrival time distributions of electrons in Das and D\"urr's works, given that all particles share their spatial states.

From these facts, it is possible to conclude that models including spin increase the explanatory power of PWT as defined in Section \ref{PWT}, whose equations are instead inadequate for the treatment of spin-$1/2$ particles. More precisely, we should say that the spin component present in the frameworks introduced above is an indispensable element for the explanation of the particles' motion in an atom's energy levels as well as for arrival times. Consequently, taking into consideration the indispensability criterion enunciated in \cite{Allori:2018}---suggesting that one should be committed only to those entities that are essential to explain physical phenomena---there are solid grounds to consider such an additional degree of freedom real.
Hence, contrary to the received view in BM, one can conclude that there is argumentative room to support the existence of spin in PWT.\footnote{Relying on the indispensability argument mentioned above, a related, cogent issue emerges forcefully: since spin is viewed in many pilot-wave models as an additional component of the wave function, and since both spin and wave functions are essential explanatory elements, then one is led to consider $\psi$ as a real entity, assigning to it serious ontological commitment---indeed, in BM particles are \emph{physically guided} by $\psi$, which has a \emph{causal} role in determining the motion of the corpuscles. Without a symmetrical explanatory work done by the wave function \emph{and} the particles, in fact, BM would not be able to explain the observed measurement outcomes. As seen in Section \ref{wf}, there are many options for a realist interpretation of $\psi$ in PWT, so it would be interesting to understand how such approaches can be modified to include spin. This topic, however, lies outside the scope of the present work and will be addressed in future research.}

\section{Extensions of Pilot-Wave Theory}
\label{Extensions}

After having discussed central topics in PWT, such as quantum measurements, contextuality, and the nature of spin, this Section reviews some of the major extensions of this framework to Quantum Field Theory (QFT) on the one hand, and to special relativity on the other. The Bohmian QFTs we shall discuss not only show how PWT can be extended beyond the realm of standard QM, but also that QFT can be given a particle ontology, contrary to those no-go theorems proving the impossibility of a corpuscular metaphysics for QFT (cf. \cite{Malament:1996}, \cite{Halvorson:2002}).\footnote{\cite{Oldofredi:2018} illustrates how Bohmian QFTs circumvent these no-go theorems.} 

Similarly, relativistic Bohmian models help us clarify what criteria make a theory genuinely relativistic and what difficulties PWT encounters, for the question of how to combine quantum nonlocality with relativity is one of the central open challenges that any quantum interpretation (including the standard account) faces. In particular, the reader will see diverse strategies to make PWT's equations invariant under Lorentz transformations---an essential symmetry that any relativistic theory should not violate. 

Lastly, the following discussion aims to show that the pilot-wave approach is not a priori incompatible with relativity: even though the frameworks we are going to introduce are initial steps toward a fully genuine relativistic PWT, they show (i) that Lorentz-invariant Bohmian models can be formulated, and (ii) that their nonlocality cannot be used to send faster-than-light signals.

\subsection{Quantum Field Theories}
\subsubsection{Stochastic Models: Bell-type Quantum Field Theory and Recent Advancements}

\cite{Durr:2005} generalizes BM to QFT, refining a stochastic model outlined in \cite{Bell:1986aa}, hence the name \emph{Bell-type QFT} (BTQFT). According to this proposal, a physical system is described by a pair $(Q_t, \Psi_t)$, where the former represents a configuration of identical particles, and the latter is the state vector which belong to an appropriate Fock space (defined as a $N$-particle Hilbert space): symmetric or anti-symmetric depending on the particles considered, bosons in the former case, fermions in the latter. 

Contrary to standard BM, the dynamical laws of BTQFT introduce variations in the number of particles to describe particle creation/annihilation, in agreement with QFT's phenomenology. These events, in turn, are represented by discontinuities in the particles' trajectories, introducing thereby a stochastic element in the model, for there is no physical process that causes them. 

As customary in PWTs, $\Psi$ evolves according to the Schr\"odinger equation:

\begin{equation}
i\hbar\frac{d\Psi_t}{dt}=H\Psi_t,
\end{equation}

\noindent{where} $H$ could be the Schr$\ddot{\mathrm{o}}$dinger or the Dirac Hamiltonian. 
As is well-known, in QFT  the Hamiltonian is a sum of terms: $H_{tot}=H_0+H_{I}$, where the first term corresponds to free processes and the second term describes the interactions. 

Between creation and annihilation events, the particles follow deterministic trajectories governed by the Bohmian law:

\begin{equation}
\frac{dQ_t}{dt}=v^{\Psi_t}(Q_t),
\end{equation}

\noindent{which} depends on the free part of $H$. The interaction Hamiltonian $H_I$, instead, represents the discontinuities of the particles' trajectories, which are represented by \emph{jump rates} $\sigma=\sigma(q',q,t)=\sigma^{\Psi_t}(q',q)$. These jumps, in turn, correspond to transition probabilities in the interval of time $(t, t+dt)$ from a given configuration of particles $q$ to another one $q'$ which differs in the particles' number.

Picture 7a below represents the emission of a photon at time $t_1$ (dashed line) from an electron and its absorption at time $t_2$ by another electron. These two events correspond to a creation and an annihilation event, respectively. The photon emission is associated with a jump with rate $\sigma(q',q,t)$: the starting configuration $q$ consists of two electrons, and the arrival configuration $qì$ also includes a photon. At time $t_2$, another jump occurs, and the particle number decreases because the photon gets absorbed. Between jumps, the particles follow deterministic trajectories. Picture 7b represents the creation of an electron-positron pair at the end of a photon trajectory. The destinations and times of the jumps are the stochastic elements of the model, and the probabilities governing these jumps are described by Markov processes: they depend only on the present state of the configuration, not on the past histories of the particles' trajectories.
\vspace{2mm}
\begin{center}
\includegraphics[scale=0.4]{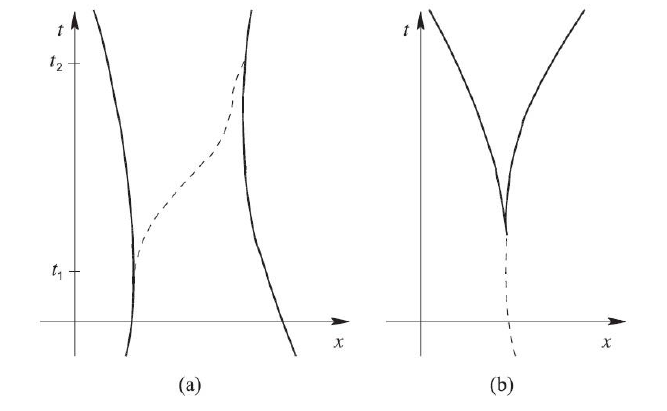}

Figure 7: Idealized representation of particle creation and annihilation events in BTQFT. The picture is taken from \cite{Durr:2004aa}.

\end{center}
 
Finally, \cite{Durr:2005} shows that BTQFT preserves equivariance: if the particle configuration $Q(t_0)$ is chosen randomly with distribution $|\Psi(t_0)|^2$, then at any later time $Q(t)$ is distributed with density $|\Psi(t)|^2$. Since both $H_0$ and $H_I$ are, by construction, associated with equivariant Markov processes, equivariance is recovered in this extension of BM. Thus, the empirical equivalence has been achieved with any \emph{regularized QFT}. 

The notions of equivariance and ``process additivity'' are the key features of BTQFT, since they are the guiding principles in the construction of the dynamics: the processes associated with $H_0$ and $H_{I}$ are defined in a manner which allows them to yield \emph{typical} histories for the primitive variables compatible with quantum statistics. Thus, it follows that BTQFT is the natural process associated with $H$ in QFT: the sums of equivariant generators for the transition probabilities define a unique equivariant process associated with sums of Hamiltonians.

Notably, the stochastic part of the dynamics resembles the processes of the wave function collapses in GRW theories, since in both cases these stochastic processes are \emph{spontaneous} in a precise sense: they are not caused or induced by external factors such as measurements, external observers, forces, etc. More precisely, in GRW theory, the evolution of the wave function is given by stochastic jump processes in Hilbert space, which are responsible for the random collapses of $\psi$. Between these random processes, the latter evolves deterministically according to the Schrödinger equation. As shown above, the BTQFT replicates this schema. Moreover, as in GRW, the BTQFT formalism provides the rates for these collapses. Thus, there is a structural similarity between the processes associated with the motion of the primitive ontologies of these theories.\footnote{For details on the similar structures of PO theories, see \cite{Allori:2008aa} and \cite{Allori:2013aa}.} 

Another stochastic extension of BM has been recently proposed by \cite{Tumulka:2025}, which provides new ideas about possible extensions of BM to Quantum Electrodynamics (QED). The author first introduces the problems affecting the standard formulation of such a theory, clarifying that some of them are inherited from non-relativistic QM, e.g., the lack of a clear metaphysical picture of reality, while others pertain to QED specifically, such as the absence of Born’s rule for electrons, positrons, and photons’ positions as well as ultraviolet divergences. 

Then, he proceeds to provide the basic elements of a Bohmian QED employing a multi-time wave function formalism\footnote{Cf.\ Section \ref{MTWF} for more details.}, implementing a particle ontology, and thereby defining trajectories for both fermions and photons depending on a preferred foliation of space-time into spacelike hypersurfaces---although Tumulka claims that satisfactory equations for photon trajectories have not been found yet. The equations of motion for such a Bohmian QED present terms governing pair creation and annihilation, which are interpreted as inherently stochastic events, as in Bell-type QFT.

\subsubsection{The Dirac sea model}

Let us now turn to a deterministic version of a Bohmian QFT based on the Dirac Sea (DS) as formulated in \cite{Deckert:2016aa}.\footnote{Bohm employed this framework in \cite{Bohm:1953aa}.}

Contrary to BTQFT, in DS there is a configuration of $N$ \emph{permanent} particles, that is, the particle number is fixed as in BM; then, a different explanation for particle creation and annihilation events is provided.

Moreover, ontological commitment is conceded to fermions only, since it is the minimal and sufficient commitment able to explain measurement outcomes (\cite{Bell:2004aa}). Bosons are not considered beables in this model, but part of its dynamical structure---this is the second major ontological difference with respect to the theories introduced above. 

\noindent To sketch the Dirac sea model, a set of assumptions is needed:
\begin{enumerate}
   \item Only the electron sector of the Standard Model is considered;
   \item Only electrodynamic interactions are included;
   \item Interactions with other particles' sectors are modeled by a time-dependent external potential\footnote{If one considers only the electrodynamic interaction among electrons, the consequence would be that they would repel each other; hence, the spatial extensions among these fermions will become larger and larger, giving rise to an unphysical behavior. To avoid this situation, one must consider an external potential that models the interaction between the rest of the particles and the set of electrons, where the motion of the other particles imposes constraints on the electrons' behavior. For details cf.\cite{Deckert:2016aa}.};
   \item The Universe is assumed to have a finite volume;
   \item The momenta of the electrons are restricted to be lower than some ultraviolet momentum cut-off $\Lambda$.
\end{enumerate}

\noindent Although these assumptions contain important simplifications, the DS model still proves able to describe the phenomena of electron-positron pair creation. 

The DS model implements two dynamical laws, one for the wave function and one for the particle configuration. Here $\Psi$ is an anti-symmetric, square-integrable, $N$-particle spinor-valued function in $\mathbb{R}^{3N}$ that evolves according to
\begin{equation}
\label{DS_WF}
i\hbar\partial_t\Psi_t(x_1,\dots,x_N)=H^N\Psi_t(x_1,\dots,x_n).
\end{equation}

\noindent The Hamiltonian $H^N$ has the particular form:

\begin{equation}
\label{DSham}
H^N=\sum^N_{k=1}\Big{(}H^0_k(x_k)+V_k(t, x_k)+H^I_k(x_k)\Big{)}
\end{equation}

\noindent whose terms are: 
\begin{itemize}
   \item The free Hamiltonian, where the $H^0(x)=-ic\alpha\cdot\nabla_x+\beta{m}c^2$. Here $\alpha, \beta$ are the $4\times4$-matrices introduced in the Dirac equation.
   \item The effective interaction of all the particles on the $k^{th}$ electron is given by the  time-dependent potential $V_k(t,x)$ for some external potential $V(t,x)$. 
   \item The last summand corresponds to the interaction Hamiltonian. The interaction among electrons is modeled by the Coulomb potential $U(x)=\frac{e^2}{4\pi\epsilon_0}|x|^{-1}$: $\epsilon_0$ is the electric constant, and $e$ represents the charge of the electron.  
\end{itemize}

The particles follow deterministic trajectories according to a guiding equation that depends on the wave function:
\begin{equation}
\label{DSvelocity}
v_t(X)=c\Big{(}\frac{j^{(k)}_t(X)}{\rho_t(X)}\Big{)}_{k=1,\dots, N}
\end{equation}
\noindent where $c$ is the speed of light, $X$ the actual configuration of $N$ electrons, the numerator is the quantum current generated by $\Psi_t$, whereas the denominator represents the the probability density generated by $\Psi_t$.

This model, as BTQFT, reproduces the statistics of the standard regularized QFT via equivariance: the form of the guiding equation ensures that if an arbitrary initial particle distribution at a certain initial time $t$ is Born-distributed, then this distribution will hold for any future time $t'$. 
\vspace{2mm}

According to \cite{Deckert:2016aa} and \cite{Esfeld:2017}, an essential feature of the DS model is its peculiar vacuum state, which is not empty---as the name may suggest---but rather filled with a homogeneous sea of particles that occupy all the negative-energy states. The Coulomb interaction is approximated in a way that the particles effectively ``ignore each other'', as if they were in a vacuum. In virtue of the Pauli exclusion principle, then, positive-energy particles do not fall into lower and lower energy states. These authors, as we have seen in Section \ref{nom}, propose a minimalist ontology of `naked' matter points, where intrinsic properties are just part of the dynamic structure; in this way, `electron', `position', `negative-state particle' become just labels to represent different patterns of motion (cf.\ \cite{Goldstein:2005b, Goldstein:2005a}). Consequently, the negative-energy states become just a tool to describe the motion of the positrons, and, therefore, they do not pose any interpretational problem in this framework.

Excitations of the vacuum state happen if an external interaction $V(t,x)\neq0$ is introduced, which includes the external influences of the other particles on the electrons. A single excitation is encoded by a two-particle wave function $\eta_t(x,y)$. Its $x$ tensor component, which we shall refer to as the \emph{electron component}, tracks the evolution of the initial excitation $\chi$; its $y$ tensor component, which we shall refer to as the \emph{hole component}, tracks the evolution of the corresponding state in the vacuum state.

Finally, let us illustrate how DS describes particle creation and annihilation events. In \cite{Deckert:2016aa}, the model is defined on an $N$-particle Hilbert space $\mathcal{H}^N$; according to this formalism, there are no particle creation and annihilation operators. However, writing the model using a Fock space, $\mathcal{F}$, one can keep track of the wave function excitations with respect to the vacuum state, allowing for the treatment of a variable number of particles. One can naturally define a Fock space via the introduction of the creation and annihilation operators. Since there is an isomorphism between the ($N$-particle sector of the) Fock space and the $N$-particle Hilbert space representations, one may recast the $N$-particle dynamics generated by the Hamiltonian \eqref{DSham} in terms of the creation and annihilation operator formalism, obtaining the canonical second-quantized Hamiltonian one encounters in QED (when neglecting radiation). 
In sum, terms like ``creation'' and ``annihilation'' here refer to a particular mathematical structure---the Fock space representation---used to describe a variable number of particles. This fact does not imply physical changes in particle number in space, which remains constant. 

\subsubsection{Field Ontologies}
\label{field}

The very first extension of PWT is given in the Appendix of  \cite{Bohm:1952ab}, where Bohm generalized his approach to electromagnetism. He provided an equation of motion for the actual field configuration $\mathrm{\textbf{A}}^T(x)$, which corresponds to the transverse part of the vector potential (more precisely, the fields generated from it), assuming a field variable as a local beable. The equation for the free quantized electromagnetic field is (following the notation of \cite{Struyve:2010aa}):
\begin{align}
\frac{\partial{\mathrm{\textbf{A}}^T(x,t)}}{\partial{t}}=\frac{\delta{S}(\mathrm{\textbf{A}}^T(x,t))}{\delta\mathrm{\textbf{A}}^T(x)}\mid_{\mathrm{\textbf{A}}^T(x)=\mathrm{\textbf{A}}^T(x,t)}\nonumber
\end{align}
\noindent{where} $S$ is the phase of the wave functional $\Psi(\mathrm{\textbf{A}}^T, t)$ which satisfies the functional Schrödinger equation, 
and the quantum equilibrium distribution is given by $|\Psi(\mathrm{\textbf{A}}^T, t)|^2\mathcal{D}\mathrm{\textbf{A}}^T$. Unfortunately, Bohm believed that this theory cannot be extended to treat fermionic fields. 

Contrary to this opinion, Valentini showed that it is possible to provide a field-theoretical characterization of fermions. One of the main motivations for such an ontology, according to Valentini, comes from the asymmetrical description of fermions and bosons in Bohm's writings on QFT, where fermions are described as particles and bosons as fields. To solve this issue, the author proposed to represent fermions as fields. As nicely expressed by Struyve, this ``approach starts from a functional Schr\"odinger representation in which the wave functionals are defined on a space of Grassmann fields (that is, anti-commuting fields). The actual configurations are Grassmann fields, which are guided by the wave functional'' (\cite{Struyve:2010aa} p.\ 5).

In this theory, the dynamical equations of motion generally describe a non-local evolution of a scalar field in 3-dimensional space, where the velocity at a specific point depends on the instantaneous values of the field at arbitrarily distant points, entailing thereby that space and time are absolute (cf.\ \cite{Valentini:1996}, p.\ 54); alternatively, the model presupposes an underlying preferred reference frame. Interestingly, in this model, quantum particles are described in a low-energy approximation in which the field mimics a standard non-relativistic wave function $\psi$: if the latter is localized in a small packet of width $\Delta x$, ``the field will be a `lump' of size $\Delta x$'' (\emph{ibid.}). 
Thus, Valentini redefines in this theory the notion of `particle' which is not a point-like object lying somewhere in space, and more importantly, is not fundamental since it emerges from a field theoretical limit. It is very interesting to note that in this pilot wave model, the particles' positions become contextual variables. Describing what happens during a position measurement, Valentini affirms that ``[t]he initial field may be completely delocalised. During the `measurement' process, the field is constrained to become concentrated in a small region around some point $x$, the `measured position'. The value of $x$ is determined by the initial field and apparatus variables (given the initial wave function and Hamiltonian).
It has no simple relation to the initial field alone. Thus, in field theory, `particle position' is a contextual variable'' (\cite{Valentini:1996}, pp.\ 54-55).\footnote{More details on Valentini's views on quantum field theory are given in \cite{Valentini:2025}; for critical remarks on his approach, see \cite{Struyve:2010aa}.}

\subsection{Approaches to Relativity}

\subsubsection{The Hyper-Surface Bohm-Dirac Model}

The non-relativistic theory introduced in Section \ref{BM} was generalized \cite{Durr:1999}---using a relativistically covariant formalism---to the case of many particles obeying the relativistic Dirac equation for electrons in external (classical) electromagnetic fields.\footnote{A similar strategy for extending Bohm's theory to the Dirac equation is given in \cite{Bohm:1993aa}.}

In this relativistic Bohmian theory, the guiding law \eqref{guide} can be written with a velocity field depending on the multi-bispinors wave functions $\Psi_{\alpha_1,\alpha_2,..., \alpha_N}(\{\mathbf{X}_k(t)\},t)$ (where $\alpha_j=0,1,2,3$ are bispinor indices). 
According to this model, the kinematic space is defined as the set of $N$-tuples of wordlines defined in Minkowski spacetime. Referring to this, it is relevant to stress that such particle trajectories are still synchronized by a common time $t$. Indeed, although this is a relativistic framework, it requires the introduction of a preferred Lorentz frame $\mathcal{R}_0$---i.e.\ a specific space-time foliation $\mathcal{F}_0$---for computing the various trajectories in time.\footnote{Cf.\ \cite{Bricmont:2016aa}, pp.\ 172-173, and \cite{Goldstein:2025}, Section 14, for an informal yet conceptually precise introduction to the main ideas introduced in this section.}  The need to introduce such a foliation derives from the fact that a Bohmian universe is uniquely described by a given particle configuration, which is constituted by the simultaneous positions of all the corpuscles.

Once this foliation is defined, one can express the particles' trajectories in a relativistically covariant way by introducing four-vectors $Q_j(s_j|\Psi,\mathcal{F}_0):=[\mathbf{X}_j(t),t]$ labelled by parameters $s_j$ growing along the trajectories (for example $s_j$ could be the proper time for particle $j$, cf.\ equations (10) and (14) in \cite{Durr:1999}). The dynamics is thus explicitly dependent on the preferred foliation $\mathcal{F}_0$ defining an absolute synchronization between the particles. In a different Lorentz frame $\mathcal{R'}$, the trajectories are obtained from the ones calculated in the preferred frame $\mathcal{R}_0$ by using Lorentz transformations. A key element of this theory is that nothing is said about which foliation $\mathcal{F}_0$ one must choose. Moreover, if another preferred foliation $\mathcal{F}'_0$ is selected, $\mathcal{F}'_0$ will in general lead to different trajectories $Q'_j(s'_j|\Psi,\mathcal{F}'_0)$.\footnote{Against this background, \cite{Galvan:2015} introduces a "no-law model" for relativistic BM that eliminates the need for a singular preferred foliation. By proposing that all spacetime foliations are equally valid and treating the universe as a union of all associated probability spaces, the model maintains empirical equivalence with QM while removing unobservable, non-Lorentz-invariant structures. In this approach, no preferred reference frame is postulated, and thereby no privileged foliation is selected.}

A second central feature of this relativistic version of BM is that the theory recovers statistical predictions of standard quantum mechanics. The subtleties concerning how to define and guarantee equivariance in this model can be found in \cite{Durr:1999}, Section III.

Building on the Hypersurface Bohm-Dirac model, \cite{DGNSZ} provided a further extension of BM to relativity. However, contrary to the strategy followed in \cite{Durr:1999}, where preferred foliations were introduced as additional, absolute spacetime structures, in this proposal, the required foliation for the particles' trajectories is embedded in the wave function, thus achieving Lorentz invariance more naturally.\footnote{As Goldstein notes, other supporters of the pilot-wave theory hold other perspectives concerning Lorentz invariance: according to \cite{Bohm:1993aa}, \cite{Holland:1993}, and \cite{Valentini:2025}, ``a detailed description of microscopic quantum processes, such as would be provided by a putative extension of Bohmian mechanics to the relativistic domain, must violate Lorentz invariance. In this view, Lorentz invariance in such a theory would be an emergent symmetry obeyed by our observations'' (\cite{Goldstein:2025}).}

\subsubsection{Multi-Time Wave-Functions}
\label{MTWF}

\cite{Lienert:2017, Tumulka:2020} provide a different way to extend BM to a Lorentz-invariant, frame-independent relativistic framework with respect to the model introduced above. 

The usual non-relativistic wave function $\psi(x_1, x_2, \dots, x_n, t)$ employed in PWT is a function of a given configuration of $N$ particles evaluated at the same time $t$, therefore, a preferred inertial frame referring to the $N$ particle positions at the {\em same} time must be selected. This approach, on the contrary, employs a {\em multi-time} function which assigns to each particle in a configuration a set of space-time coordinates:

\begin{align}
\phi\bigg(  (t_1, x_1), (t_2, x_2), \dots, (t_n, x_n) \bigg).
\end{align}

\noindent In virtue of such individual assignment, the $\phi$-function does not require a choice of a preferred frame. In the special case in which $t_1 = t_2 = \dots = t_n$, one recovers the usual non-relativistic wave function, which is then just a special case of a multi-time function---or, as Lienert et al put it, ``$\psi$ is the restriction of $\phi$ to the simultaneous configurations relative to the chosen reference frame''. Notably, $\phi$ is a covariant ``particle-position representation of the state vector'' (\cite{Tumulka:2018}; cf.\ \cite{Lienert:2017}, p.\ 2 for details).

The authors then proceed to specify the law of motion for $\phi$ from initial conditions, which, in this framework, refers to the values of $\phi$ at those configurations for which all $t_j =0$, whereas the particles' positions are arbitrary. A multi-time wave function defines a system of linear partial differential equations, called {\em multi-time Schr\"odinger equations}, with one equation for each time coordinate (with $\hbar=1$):

\begin{align}
i\frac{\partial \phi}{\partial t_j}=H_j\phi, \;\;\;\;\; j=1, \dots, N.
\end{align}

\noindent This set of dynamical equations is Lorentz-invariant since it treats spatial and temporal coordinates symmetrically, contrary to \eqref{SE}, which embeds a first-order derivative in time, but a second-order derivative for space. Moreover, to guarantee the preservation of symmetry in the treatment of space and time coordinates, the laws of motion for $\phi$ implement relativistic wave operators (e.g., Dirac or Klein-Gordon operators).\footnote{To fully formulate the dynamics of the multi-time approach, we need to mention the coonsistency condition, $[i\partial_{t_j}-H_j, i\partial_{t_k}-H_k]=0$, $\forall j\neq k$. This is necessary and sufficient to guarantee the existence of a joint solution for all the initial conditions. For details about the justification of such a condition, cf.\ \cite{Lienert:2017, Tumulka:2020}.} As usual in a Bohmian setting, the particles are guided by the phase and amplitude of the $\phi$-function; however, to define particles' trajectories, one needs to introduce a spacetime foliation, for the velocity of one particle depends on the instantaneous position of the other $N-1$ corpuscles evaluated at the same time $t$. As \cite{Tumulka:2018} puts it, ``The upshot is that the evolution of the wave function can be defined in a covariant way without using the time foliation $\mathcal{F}$, which then needs to be introduced for the trajectories''. 

\section{Conclusions and Outlook}
\label{conc}

This Chapter introduced the main formulations of PWT and discussed relevant philosophical issues concerning the status of Born's rule and the nature of $\psi$. In addition, the pilot-wave account of quantum measurement has been reviewed. Building on such measurement theory, I argued that PWT is simpler than standard QM for two reasons: on the one hand, the collapse rule becomes a theorem of the pilot-wave dynamics; hence, the number of postulates {\em decreases} in the causal approach. On the other hand, using a historical case study of Dirac's analogy between classical and quantum theories, I showed that the mathematical machinery of PWT is deeply embedded in the quantum-mechanical formalism. Thereby, PWT does not add structures to standard QM. 

Furthermore, I reviewed the argument used by contemporary supporters of BM to rule out the existence of spin. Contrary to this view, I suggested that an ontological interpretation of spin can be fruitful, for it increases the explanatory power of PWT. More precisely, I share the claim that spin is not a genuine property instantiated by Bohmian particles, so that the spin observable, qua inherent attribute of the particle, does not refer to any actual metaphysical quality. However, the reduction of spin measurements to position measurements is not sufficient to preclude the possibility of defining spin in a different way with respect to standard QM and to include this new characterization of it in PWT, assigning it ontological commitment.
Indeed, as we have seen, spin can be formalized as an additional degree of freedom influencing the particles' motion. In these new clothes, spin becomes an indispensable element in explaining the particles' orbits within an atom's energy levels, as well as in computing trajectories of arrival times. Consequently, employing Allori's indispensability criterion, there is argumentative room to consider spin real in PWT. 

Finally, we reviewed and briefly discussed some compelling extensions of PWT to QFT and relativity theory. In this regard, before concluding, I want to draw the reader's attention to the various Bohmian approaches to quantum gravity and cosmology. Notable examples are given by research on shape dynamics (, \cite{Durr:2020}, \cite{Vassallo:2024}, \cite{Vassallo:2025, Vassallo:2025b}), and Bohmian theories of spacetime (\cite{Goldstein:2001aa}, \cite{Neto:2005}, \cite{Neto:2019}, \cite{Valentini:2025}). These frameworks show not only that PWT is a growing and healthy research program but also that the central foundational questions of contemporary physics are at the center of the agenda in the pilot-wave community, looking for new directions towards a unification between the most fundamental theories of spacetime and a metaphysically clear account of the quantum.

\bibliographystyle{apalike}
\bibliography{PhDthesis}

\section*{Further Readings}

\noindent - Allori, V. (2013). Primitive ontology and the structure of fundamental physical theories. In Albert, D.Z. and Ney, A. (eds.) {\em The Wave Function: Essays on the Metaphysics of Quantum Mechanics}, pages 58–75. New York: Oxford University Press.
\vspace{2mm}

\noindent - Beller, M. (1999). {\em Quantum Dialogue: The Making of a Revolution}. Chicago: University of Chicago Press.
\vspace{2mm}

\noindent - Bricmont, J., Goldstein, S. and Hemmick, D. (2022). From EPR-Schrödinger Paradox to Nonlocality Based on Perfect Correlations, {\em Foundations of Physics}, 52.
\vspace{2mm}

\noindent - Cushing, J. (1994). {\em Quantum Mechanics: Historical Contingency and the Copenhagen Hegemony}. Chicago: University of Chicago Press.
\vspace{2mm}

\noindent - Cushing, J., Fine, A. and Goldstein, S. (eds.) (1996). {\em Bohmian Mechanics and Quantum Theory: An Appraisal}. Dordrecht: Springer.
\vspace{2mm}

\noindent - Daumer, M., Dürr, D., Goldstein, S. and Zanghì, N. (1997). Naive Realism About Operators, {\em Erkenntnis}, 45(2):379–397.
\vspace{2mm}

\noindent - Dewdney, C. (2023). Rekindling of de Broglie–Bohm Pilot Wave Theory in the Late
Twentieth Century: A Personal Account. {\em Foundations of Physics}, 53, 24.
\vspace{2mm}

\noindent - Dürr, D. and Lazarovici, D. (2020). { \em Understanding Quantum Mechanics: The World According to Modern Quantum Foundations}. Cham: Springer.
\vspace{2mm}

\noindent - Freire Junior, O. (2015). {\em The Quantum Dissidents. Rebuilding the Foundations of Quantum Mechanics (1950-1990)}. Berlin: Springer.
\vspace{2mm}

\noindent - Hemmick, D. L. and Shakur, A.M. (2012). {\em Bell’s Theorem and Quantum Realism: Reassessment in Light of the Schrödinger Paradox}. New York: Springer.
\vspace{2mm}

\noindent - Hiley, B. (2005). Non-Commutative Quantum Geometry: A Reappraisal of the Bohm
Approach to Quantum Theory. In Elitzur, A.C., Dolev, S. and Kolenda, N. (eds)
{\em Quo Vadis Quantum Mechanics?}. Berlin: Springer.
\vspace{2mm}

\noindent - Leavens, C.R. (2008). Bohm Trajectory Approach to Timing Electrons. In Muga, J.G., Sala Mayato, R. and Egusquiza, I.L. (eds.) {\em Time in Quantum Mechanics}. Berlin-Heidelberg: Springer.
\vspace{2mm}

\noindent - Lewis, P.J. (2007). Empty Waves in Bohmian Quantum Mechanics, {\em British Journal for the Philosophy of Science}, 58(4): 787–803.
\vspace{2mm}

\noindent - Maudlin, T. (1994). {\em Quantum Non-Locality and Relativity: Metaphysical Intimations of Modern Physics}. Cambridge, MA: Blackwell.
\vspace{2mm}

\noindent - Maudlin, T. (2019). {\em Philosophy of Physics:\ Quantum Theory}. Princeton: Princeton University Press.
\vspace{2mm}

\noindent - Oldofredi, A. (2025). {\em Guiding Waves in Quantum Mechanics. One Hundred Years of de Broglie-Bohm Pilot-Wave Theory}. Oxford: Oxford University Press.
\vspace{2mm}

\noindent - Oriols, X. and Mompart, J. (2019). {\em Applied Bohmian Mechanics}. Singapore: Pan Stanford Publishing.
\vspace{2mm}

\noindent - Tumulka, R. (2021). Bohmian Mechanics, in Knox, E. and Wilson, A. (eds), {\em The Routledge Companion to the Philosophy of Physics}. Routledge.
\vspace{2mm}

\noindent - Valentini, A. {\em Beyond the Quantum. A Quest for the Origin and Hidden Meaning of Quantum Mechanics}. Oxford: Oxford University Press.
\vspace{2mm}

\noindent - Vassallo, A. and Esfeld, M. (2013). A Proposal for a Bohmian Ontology of Quantum
Gravity. {\em Foundations of Physics}, 44(1): 1–18
\vspace{2mm}

\end{document}